\documentclass[journalpaper,10pt,twocolumn]{IEEEtran}

\usepackage{amssymb}
\usepackage{graphicx}
\usepackage{epstopdf}
\usepackage{hyperref}
\usepackage{multirow}
\usepackage[caption=false]{subfig}
\usepackage{tabularx}
\usepackage{amsmath}
\usepackage[english]{babel}
\usepackage{algorithm}
\usepackage{algpseudocode}
\usepackage{algpseudocode,algorithmicx}
\usepackage{pifont}
\usepackage{enumitem}
\usepackage{booktabs}
\usepackage{makecell}
\usepackage{lineno,hyperref}
\usepackage{mathtools,nccmath}
\usepackage{etoolbox}
\usepackage{tikz}
\usetikzlibrary{tikzmark}
\usetikzlibrary{calc}
\modulolinenumbers[5]

\hypersetup{
    colorlinks,%
    citecolor=black,%
    filecolor=black,%
    linkcolor=black,%
    urlcolor=black,
    bookmarks=false
}

\addto\captionsenglish{\renewcommand*{\tablename}{TABLE}}
\addto\captionsenglish{\renewcommand*{\figurename}{Fig.}}

\DeclareMathOperator{\E}{\mathbb{E}}

\errorcontextlines\maxdimen

\newcommand{\ALGtikzmarkcolor}{black}
\newcommand{\ALGtikzmarkextraindent}{4pt}
\newcommand{\ALGtikzmarkverticaloffsetstart}{-.5ex}
\newcommand{\ALGtikzmarkverticaloffsetend}{-.5ex}
\makeatletter
\newcounter{ALG@tikzmark@tempcnta}

\newcommand\ALG@tikzmark@start{%
    \global\let\ALG@tikzmark@last\ALG@tikzmark@starttext%
    \expandafter\edef\csname ALG@tikzmark@\theALG@nested\endcsname{\theALG@tikzmark@tempcnta}%
    \tikzmark{ALG@tikzmark@start@\csname ALG@tikzmark@\theALG@nested\endcsname}%
    \addtocounter{ALG@tikzmark@tempcnta}{1}%
}

\def\ALG@tikzmark@starttext{start}
\newcommand\ALG@tikzmark@end{%
    \ifx\ALG@tikzmark@last\ALG@tikzmark@starttext
    \else
        \tikzmark{ALG@tikzmark@end@\csname ALG@tikzmark@\theALG@nested\endcsname}%
        \tikz[overlay,remember picture] \draw[\ALGtikzmarkcolor] let \p{S}=($(pic cs:ALG@tikzmark@start@\csname ALG@tikzmark@\theALG@nested\endcsname)+(\ALGtikzmarkextraindent,\ALGtikzmarkverticaloffsetstart)$), \p{E}=($(pic cs:ALG@tikzmark@end@\csname ALG@tikzmark@\theALG@nested\endcsname)+(\ALGtikzmarkextraindent,\ALGtikzmarkverticaloffsetend)$) in (\x{S},\y{S})--(\x{S},\y{E});%
    \fi
    \gdef\ALG@tikzmark@last{end}%
}

\apptocmd{\ALG@beginblock}{\ALG@tikzmark@start}{}{\errmessage{failed to patch}}
\pretocmd{\ALG@endblock}{\ALG@tikzmark@end}{}{\errmessage{failed to patch}}
\makeatother

\begin{document}

\title{Meta-learning with Latent Space Clustering in Generative Adversarial Network for Speaker Diarization}

\author{Monisankha~Pal,~\IEEEmembership{Member,~IEEE,}
        Manoj~Kumar,~\IEEEmembership{Member,~IEEE,}
        Raghuveer~Peri,~\IEEEmembership{Member,~IEEE,}
        Tae Jin~Park,~\IEEEmembership{Member,~IEEE,}
        So Hyun~Kim,
        Catherine~Lord,
        Somer~Bishop,
        and~Shrikanth~Narayanan,~\IEEEmembership{Fellow,~IEEE}
\thanks{M. Pal, M. Kumar, R. Peri, T. J. Park and S. Narayanan are with Signal Analysis and Interpretation Laboratory, University of Southern California, Los Angeles, USA e-mail: (mp$\_$323@usc.edu; prabakar@usc.edu; rperi@usc.edu; taejinpa@usc.edu; shri@sipi.usc.edu).}
\thanks{S. H. Kim is with Center for Autism and the Developing Brain, Weill Cornell Medicine, USA e-mail:(sok2015@med.cornell.edu)}
\thanks{C. Lord is with Semel Institute of Neuroscience and Human Behavior, University of California Los Angeles, USA e-mail:(CLord@mednet.ucla.edu)}
\thanks{S. Bishop is with Department of Psychiatry, University of California, San Francisco, USA e-mail:(somer.bishop@ucsf.edu)}}


\maketitle

\begin{abstract}
The performance of most speaker diarization systems with x-vector embeddings is both vulnerable to noisy environments and lacks domain robustness. Earlier work on speaker diarization using generative adversarial network (GAN) with an encoder network (ClusterGAN) to project input x-vectors into a latent space has shown promising performance on meeting data.
In this paper, we extend the ClusterGAN network to improve diarization robustness and enable rapid generalization across various challenging domains. 
To this end, we fetch the pre-trained encoder from the ClusterGAN and fine-tune it by using prototypical loss (meta-ClusterGAN or MCGAN) under the meta-learning paradigm. Experiments are conducted on CALLHOME telephonic conversations, AMI meeting data, DIHARD II (dev set) which includes challenging multi-domain corpus, and two child-clinician interaction corpora (ADOS, BOSCC) related to the autism spectrum disorder domain. Extensive analyses of the experimental data are done to investigate the effectiveness of the proposed ClusterGAN and MCGAN embeddings over x-vectors. The results show that the proposed embeddings with normalized maximum eigengap spectral clustering (NME-SC) back-end consistently outperform Kaldi state-of-the-art x-vector diarization system. Finally, we employ embedding fusion with x-vectors to provide further improvement in diarization performance. 
We achieve a relative diarization error rate (DER) improvement of 6.67\% to 53.93\% on the aforementioned datasets using the proposed fused embeddings over x-vectors. 
Besides, the MCGAN embeddings provide better performance in the number of speakers estimation and short speech segment diarization as compared to x-vectors and ClusterGAN in telephonic data.
\end{abstract}


\IEEEpeerreviewmaketitle

\vspace{-10pt}

\section{Introduction}

\IEEEPARstart{S}{}peaker diarization \cite{anguera2012speaker}, the task of determining ``who spoke when'' in a multi-speaker audio stream has a wide range of applications such as information retrieval, speaker-based indexing, meeting annotations, and conversation analysis \cite{vijayasenan2009information}. 
Present-day diarization systems typically comprise four components: (a) A speech segmentation module that removes the non-speech parts using a speech activity detector (SAD) and segments the speech part into multiple speaker-homogeneous short segments \cite{garcia2017speaker}; (b) A speaker representation (embedding) extractor that maps the segments into fixed-dimensional \emph{speaker embeddings} such as i-vectors \cite{shum2013unsupervised, senoussaoui2014study}, d-vectors \cite{wang2018speaker, zhang2019fully} and \emph{x-vectors} \cite{garcia2017speaker, sell2018diarization}; (c) A clustering module that determines the number of constituent speakers in an audio recording and clusters the extracted embeddings into these speakers \cite{park2019auto, lin2019lstm}; (d) A re-segmentation module that refines the clustering results \cite{garcia2017speaker}.
\par
For embedding extraction, typically i-vectors have been obtained through total variability space projection \cite{dehak2010front}. However, recently significant performance improvement has been shown using deep neural network embeddings such as d-vectors with architectures such as 
LSTM \cite{wang2018speaker, wan2018generalized}, CNN \cite{zajic2017speaker}; and x-vectors with time-delay neural network (TDNN) \cite{garcia2017speaker, snyder2018x}. The combination of different embeddings, e.g., c-vectors using 2D self-attentive structure, has also been proposed to exploit the complementary merits of each embedding \cite{sun2019speaker}. 
\par
In terms of clustering, most of the existing algorithms that have been used in speaker diarization are unsupervised. Among them, agglomerative hierarchical clustering (AHC) \cite{garcia2017speaker} and spectral clustering (SC) \cite{ning2006spectral} using pairwise embedding similarity measurement techniques like cosine distance \cite{wang2018speaker, park2019auto}, PLDA \cite{park2019second} and using an LSTM \cite{lin2019lstm} are the most popular. Similarly, other unsupervised clustering methods such as Gaussian mixture model \cite{shum2013unsupervised, zajic2017speaker}, mean-shift \cite{senoussaoui2014study}, k-means \cite{dimitriadis2019enhancements}, and links \cite{mansfield2018links} have also been adopted for speaker diarization. Moreover, clustering depends on tuning hyperparameters like stopping threshold (for AHC), the $p$-value for binarization of affinity matrix (for SC). However, more recently, an auto-tuning and improved version of the spectral clustering approach on x-vectors using cosine similarity measure, which is called as \emph{normalized maximum eigengap spectral clustering (NME-SC)} was introduced in \cite{park2019auto}. Despite the success of these speaker clustering algorithms, speaker diarization remains a challenging task due to the wide heterogeneity and variability of audio data recorded in many real-world scenarios \cite{Han2008Strategiestoimprovethe}.
\par
The other approach for speaker clustering has been based on supervised methods. A fully supervised speaker diarization framework, named UIS-RNN was proposed in \cite{zhang2019fully}. 
Although this model for clustering produces excellent performance in telephone conversations, its performance deteriorates in a more challenging multi-domain database like DIHARD II \cite{fini2020supervised}. 
To improve the UIS-RNN diarization performance further, a novel sample-mean loss function to train the RNN has been introduced very recently \cite{fini2020supervised}. 
Efforts have been made to automatically deal with speaker-overlapping speech segments and directly optimize an end-to-end neural network based on diarization errors \cite{Fujita2019}. 
The network is trained in a supervised manner using a permutation free objective function. The diarization performance was further enhanced by introducing a self-attention based end-to-end neural network \cite{fujita2019end}. Although the above methods do not rely on clustering and can directly compute the final diarization outputs using a single network, they assume that the number of speakers is known apriori or at least bounded to two speakers.  
Along these lines, the performance of deep embedded clustering, which was originally proposed in \cite{xie2016unsupervised}, was incorporated and modified for speaker clustering in diarization task \cite{dimitriadis2019enhancements}. 
The limitation of this work is that a good estimate of the number of speakers is needed for its evaluation.
\par
While performance of tasks such as speech and speaker recognition have improved significantly due to supervised deep learning approaches, most of the speaker clustering is yet to take advantage of similar techniques. The main problem that hinders in making 
clustering a supervised task is associated with the fact that speaker labels are ambiguous (e.g., both ``112233" and ``223311" sequences of labels are equally correct for the same diarization session). 
In our earlier proposed work, we incorporated \emph{ClusterGAN} to non-linearly transform DNN-based speaker embeddings into a low-dimensional latent space better suited for clustering \cite{pal2020speaker}. The proposed ClusterGAN, which exploits the GAN latent space with the help of an encoder network, was trained with a combination of adversarial loss, latent variable recovery loss, and clustering-specific loss. 
Although the proposed system showed significant performance improvement over x-vector based state-of-the-art in meeting and child-adult interaction corpora, its performance was not tested against telephone conversations and a broader set of multi-domain data. 
\par
In this work, a ClusterGAN network which was originally proposed for image clustering \cite{mukherjee2019clustergan}, is adopted and modified for the speaker clustering task in the speaker diarization framework.
The GAN and the encoder network are trained jointly in a supervised manner with clustering-specific loss and latent embeddings are extracted using the trained encoder to perform unsupervised clustering at the back-end. Two main advantages of GAN-based latent space clustering are the interpretability and interpolation in the latent space \cite{mukherjee2019clustergan}. We use ClusterGAN-trained encoder network as initialization to further fine-tune it with meta-learning based \emph{prototypical loss} function \cite{snell2017prototypical, wang2019centroid}. This is represented as \emph{meta-ClusterGAN or MCGAN} in this paper. 
The prototypical network was introduced for the few-shot image classification task \cite{snell2017prototypical} and is the state-of-the-art approach on a few-shot image classification benchmark. The motivation behind using proto-learning for our task is that it has a simpler inductive bias in the form of speaker prototypes and can perform rapid generalization to new speakers or types of data not seen while training. The prototypical loss trained for learning a metric space to mimic the test scenario will be beneficial in capturing information related to both generalization and clustering objectives.
\par
The main contributions of this paper are: (a) A novel speaker diarization framework based on prototypical learning; (b) Extensive multi-domain experimental evaluation and analysis of the proposed diarization system on various challenging speaker diarization corpora; (c) Demonstration of the use of novel speaker embeddings that outperform x-vectors through analysis across various challenging scenarios. 
\par

\begin{figure*}[!t]
\centering
\includegraphics[width= \textwidth]{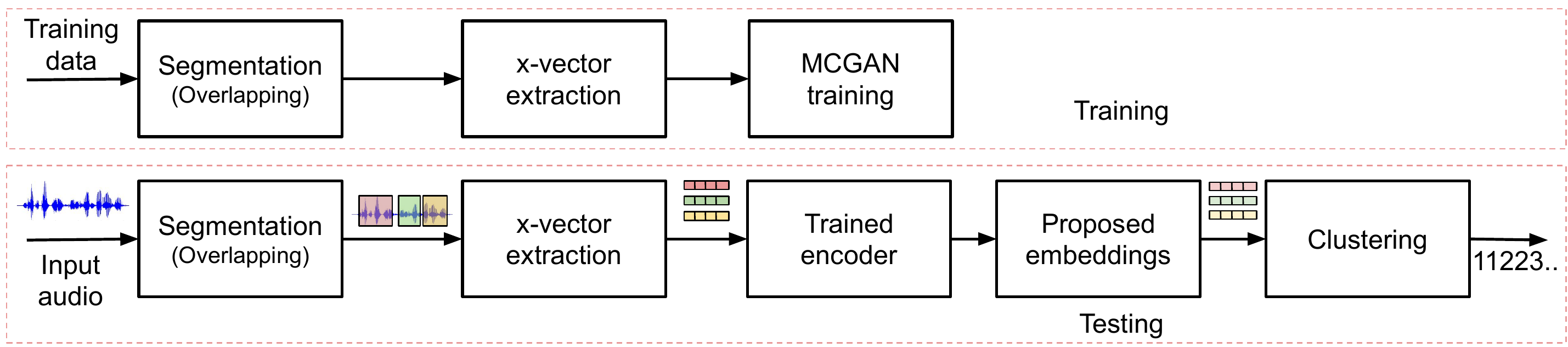}
\vspace{-18pt} 
\caption{Skeleton of the proposed speaker diarization system.}\label{fig1}
\vspace{-12pt}
\end{figure*}

\vspace{-12pt}
\section{Related works}\label{section2}

\subsection{Deep clustering algorithms}

Using deep neural networks to non-linearly transform the input data into cluster-friendly representation along with dimension reduction is commonly known as deep clustering \cite{aljalbout2018clustering}. 
Recent deep clustering methods on image data using autoencoder networks like deep embedded clustering (DEC) \cite{xie2016unsupervised}
achieve impressive clustering performance. Generative modeling based approaches like variational deep embedding \cite{jiang2016variational}, information maximizing GAN (InfoGAN) \cite{chen2016infogan}, GAN mixture model \cite{yu2018mixture} learn latent representation space and can interpolate to generate new samples from the data distribution. In all these algorithms, the deep neural network is usually trained on two types of losses: representation loss or network loss and clustering-specific loss. The network loss is essential for network initialization and is used to learn feasible latent features. The different network losses are reconstruction loss of autoencoder, variational loss of a variational autoencoder, and adversarial loss of GANs. On the other hand, clustering-specific loss helps to learn representations suitable for clustering. The option for clustering-specific losses are assignment losses like k-means loss \cite{yang2017towards}, cluster assignment hardening loss \cite{xie2016unsupervised}, agglomerative clustering loss \cite{yang2016joint}, spectral clustering loss \cite{shaham2018spectralnet} or regularization losses such as locality preserving loss, cluster classification loss \cite{aljalbout2018clustering}. Different ClusterGAN proposed for image data clustering adopts adversarial loss in GAN and clustering-specific loss like balanced self-paced entropy minimization loss \cite{ghasedi2019balanced} or cluster classification loss \cite{mukherjee2019clustergan}. Very recently, few deep clustering approaches like transformer-based discriminative neural clustering model \cite{li2019discriminative}, deep clustering loss in end-to-end neural speaker diarization \cite{fujita2019end}, deep embedded clustering \cite{dimitriadis2019enhancements}, and ClusterGAN \cite{pal2020speaker} have been used for speaker diarization. Although multifarious deep clustering approaches have been successfully applied for image data clustering, their application toward speaker diarization has been limited mainly due to 
the problem of the unknown number of speakers in a given diarization session.

\vspace{-12pt}
\subsection{Meta-learning algorithms}

Inspired by human learning of new categories (classes) given just a very few examples, the meta-learning model trained over a large variety of learning tasks can adapt or generalize well to potentially unseen tasks \cite{ravi2016optimization}. It is also known as learning-to-learn, which learns on a given task and also across tasks. In the computer vision literature, there are three common approaches to meta-learning: metric-based \cite{chen2011learning}, model-based \cite{santoro2016meta}, and optimization-based \cite{vinyals2016matching}. Metric learning aims at learning a metric or distance function over the embedding space. Among metric-learning based approaches, Siamese networks \cite{koch2015siamese} and triplet networks \cite{schroff2015facenet} for learning speaker embeddings have been proposed for speaker recognition \cite{chung2020defence} and speaker diarization \cite{bredin2017tristounet}, and have yielded promising performances. 
The prototypical network that learns a metric space by computing prototype representation of each class is a state-of-the-art approach for a few-shot image classification tasks \cite{snell2017prototypical}. Along these lines, prototypical loss to optimize a speaker embedding model for the speaker verification task was explored in \cite{wang2019centroid, chung2020defence}. The resulting model provides superior performance to triplet loss based models. Very recently, the usage of protonets for child-adult audio classification task was explored in \cite{koluguri2020meta}. Our proposed approach uses prototypical loss (PTL) to fine-tune the encoder of ClusterGAN for robust speaker embedding extraction in the speaker diarization framework.

\vspace{-12pt}
\section{Proposed speaker diarization system}\label{section3}

An overview of our proposed speaker diarization system is shown in Fig. \ref{fig1}. The non-speech part in a given multi-speaker conversation is removed first by using a speech activity detection (SAD) system. Our diarization system uses Kaldi\footnote{\url{https://kaldi-asr.org/}} style uniform segmentation and the segments are embedded into a fixed-dimensional vector using a time-delay neural network (TDNN), which is commonly known as x-vector \cite{snyder2018x}. The proposed meta-ClusterGAN (MCGAN) is developed on top of x-vectors to perform deep latent space clustering for speaker diarization. We describe each of the modules in the diarization pipeline below.

\vspace{-12pt}
\subsection{Segmentation}

In this paper, our proposed system uses oracle SAD for all the analysis and experiments, following common practice in the speaker diarization literature \cite{garcia2017speaker, zhang2019fully, diez2019analysis}. Therefore, our approach starts with a temporal uniform segmentation of 1.5 sec with an overlap of 1 sec between two adjacent segments. This denser segmentation gives more number of samples while evaluating a diarization session and it helps in clustering.

\vspace{-12pt}
\subsection{Speaker embedding vector}

The speaker embedding vectors used to train the ClusterGAN models are x-vectors, which are fixed-length representation using a TDNN from variable-length utterances. In this approach, MFCCs are first extracted at frame-level and input to a TDNN for supervised training using the categorical cross-entropy loss based on the speaker labels. The statistics pooling layer inside the TDNN architecture is used to convert frame-level features into a segment-level embedding. The detailed procedure of x-vector extraction is concisely described in \cite{garcia2017speaker, snyder2018x}. In this paper, we use Kaldi-based pre-trained x-vectors.


\vspace{-12pt}
\subsection{Meta-ClusterGAN (MCGAN) training}

\begin{figure*}[!t]
\centering
    \includegraphics[width= \textwidth]{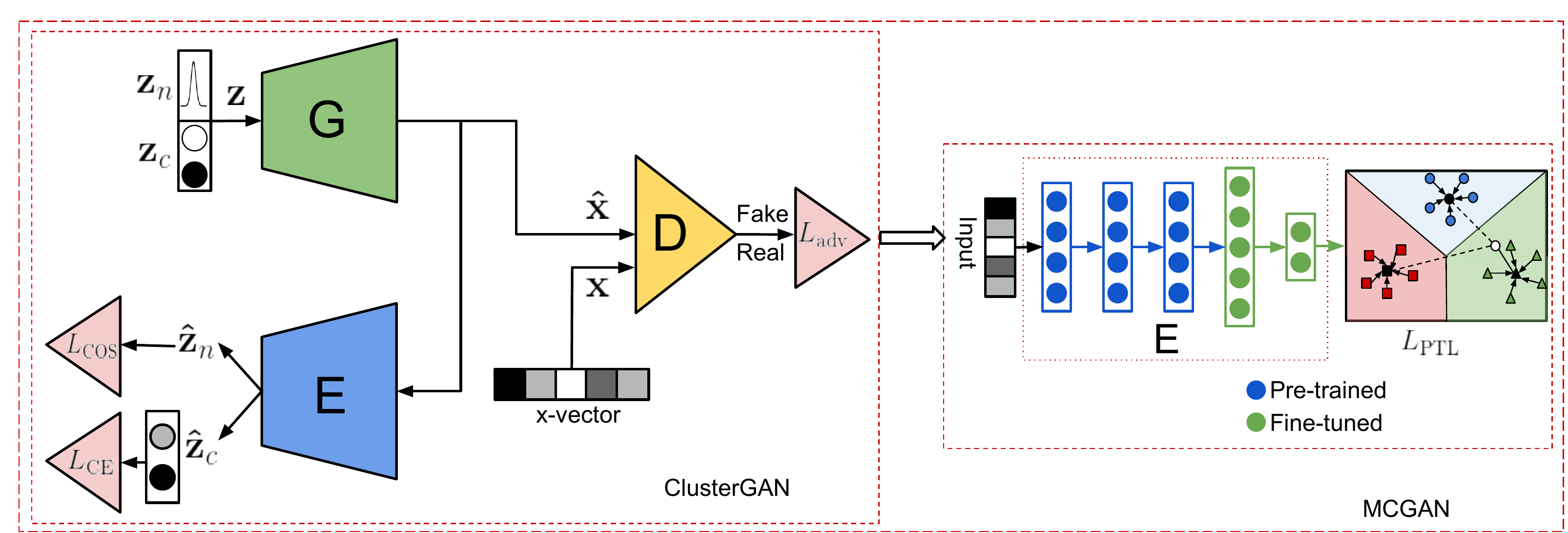}
    \vspace{-15pt} 
    \caption{MCGAN architecture. Here, $L_{\textnormal{adv}}$, $L_{\textnormal{COS}}$, $L_{\textnormal{CE}}$ and $L_{\textnormal{PTL}}$ represent adversarial, cosine distance, cross-entropy and prototypical loss functions.}\label{fig2}
\vspace{-12pt}
\end{figure*}

We employ meta-ClusterGAN (MCGAN) for the speaker embedding extraction in the speaker diarization framework. The motivation behind introducing MCGAN is to non-linearly transform the input x-vectors (trained with categorical cross-entropy loss) into another embedding 
suitable for speaker clustering and that can generalize well to new classes (here, speakers) not seen while training. As shown in Fig.~\ref{fig2}, the proposed MCGAN training has two phases: (a) parameter initialization using a ClusterGAN, trained with clustering-specific loss in GAN latent space, and (b) inducing robustness to the initialized encoder in ClusterGAN by further fine-tuning it with meta-learning based prototypical loss. We discuss each phase of training further in the following subsections.

\subsubsection{ClusterGAN training}

We pre-train the MCGAN encoder using ClusterGAN training since it can decipher the original data representation by exploiting the GAN latent space. 
The learned encoder in ClusterGAN can generate embeddings in another space while maintaining the separable properties among the classes. ClusterGAN comprises three components: generator ($G$), discriminator ($D$) and encoder ($E$). The complete ClusterGAN architecture details and its training procedure are described in detail below.

\paragraph{\textbf{Adversarial training}}

ClusterGAN adopts adversarial training of GANs for the clustering task. The standard GAN is formulated as an adversarial mini-max game between two neural networks: a generator ($G$) and a discriminator ($D$) \cite{goodfellow2014generative}. The generator aims to create a map from latent space to data space, i.e., $G: \mathbf{z} \rightarrow \hat{\mathbf{x}}$. It takes random noise $\mathbf{z}$ sampled from $p_\mathbf{z}$ and synthesizes data similar to original data to fool the discriminator. The discriminator is considered to be a mapping from the data space to a real value $D: \mathbf{x} \rightarrow \mathbb{R}$. It takes real data $\mathbf{x}$ sampled from $p_\mathbf{x}^r$ and aims to distinguish the real data from the generator produced samples. Although GANs can learn to mimic any data distribution, they are difficult to train due to the mode collapse problem \cite{arjovsky2017wasserstein}. To address this issue, several variants of GANs such as Wasserstein GAN (WGAN) \cite{arjovsky2017wasserstein}, and improved WGAN \cite{gulrajani2017improved} (IWGAN) have been proposed in the literature. In this work, we incorporate IWGAN as our GAN network. The objective function of this adversarial game between $G$ and $D$ is

\begin{equation}
\footnotesize
   \underset{G}{\mathrm{min}} \hspace{2pt} \underset{D} {\mathrm{max}} \hspace{2pt} U_{\mathrm{IWGAN}}(D, G) = \E_{\mathbf{x} \sim p_\mathbf{x}^r} \left[D(\mathbf{x}) \right] - \E_{\mathbf{z} \sim p_\mathbf{z}} \left[D(G(\mathbf{z}))\right] + \lambda \cdot \mathrm{GP}
\end{equation}

where $\lambda$ denotes the gradient penalty coefficient and $\textnormal{GP}$ represents the gradient penalty term \cite{gulrajani2017improved}. The gradient penalty term can be expressed as 
\begin{equation}
    \textnormal{GP} = \E_{\hat{\mathbf{x}} \sim p_{\hat{\mathbf{x}}}} \left[\left(\lVert \triangledown_{\hat{\mathbf{x}}} D(\hat{\mathbf{x}})\rVert _{2} - 1\right)^2\right]
\end{equation}
where $\hat{\mathbf{x}} = \epsilon \mathbf{x} + (1-\epsilon) G(\mathbf{z})$ and $\epsilon$ is a random number uniformly sampled in between 0 and 1.

\paragraph{\textbf{Mixture of discrete and continuous latent variables}} 

One possible way to perform clustering in the latent space is to back-project the data into the GAN latent space and then cluster it. The latent vectors for GANs trained with different priors such as Gaussian or uniform distribution usually lead to bad clustering \cite{lipton2017precise}. Although the latent space may contain useful information about the data, the distance geometry does not reflect any form of clustering. To combat this issue, boosting the latent space using categorical variables ($\mathbf{z}_c$) to form non-smooth geometry is essential. The discrete variable $\mathbf{z}_c$ as a mixture with the continuous random variable ($\mathbf{z}_n$) will restrict the GAN generator to produce each mode only generating samples from a corresponding category in the real data. A similar type of latent variable structure within a GAN generator for learning disentangled and meaningful representation was employed in InfoGAN \cite{chen2016infogan}. However, ClusterGAN has been reported to be superior to InfoGAN for clustering \cite{mukherjee2019clustergan}. Furthermore, continuity in the latent space is also required for good interpolation objective and GANs have good interpolation ability. Therefore, our latent variable $\mathbf{z}$ is a concatenation of $\mathbf{z}_n$ and $\mathbf{z}_c$. In this work, we use $\mathbf{z}_n \sim \mathcal{N}(0, \sigma^{2} \mathbf{I}_{d_{n}})$, where we chose a small value of variance ($\sigma$) as 0.10 to make the clusters separated. We use $\mathbf{z}_c$ as a one-hot encoded vector by using the original speaker labels in the training data. Thus, our ClusterGAN training is supervised in nature. The mixture of $\mathbf{z}_n$ and $\mathbf{z}_c$ as the prior enables clustering in the latent space.

\paragraph{\textbf{Inverse mapping network}}

Inverse mapping from data space to latent space is a non-trivial problem, as it requires the inversion of the generator, which is a multi-layered non-linear model. The work proposed in \cite{lipton2017precise, creswell2018inverting}, tackles this issue by solving an optimization problem in $\mathbf{z}$ to recover the latent vectors using $\mathbf{z}^{*} = \textnormal{argmin}_{\mathbf{z}} L(G(\mathbf{z}), \mathbf{x}) + \lambda \lVert \mathbf{z} \rVert _{p}$, where $L$ is a suitable loss function, $\lambda$ is a regularization constant and $\lVert \cdot \rVert _{p}$ denotes the norm. However, this optimization is non-convex in $\mathbf{z}$ and there exist multiple $\mathbf{z}$ values to describe a single real data $\mathbf{x}$ \cite{mukherjee2019clustergan, creswell2018inverting}. To mitigate these issues, the stochastic clipping of $\mathbf{z}$ at each iteration step was used in \cite{lipton2017precise}. However, the above approaches are not amenable to clustering. In this work, we train a separate encoder ($E$) network alongside the GAN network to learn the inverse mapping function of the generator, estimating discriminative latent embeddings for the real data. We fix $\mathbf{z}_c$ and use multiple restarts, each time sampling $\mathbf{z}_n$ from a normal distribution. Moreover, to enforce precise recovery of $\mathbf{z}_n$, we compute the numerical difference between $\mathbf{z}_n$ and corresponding encoder output $\hat{\mathbf{z}}_n$. We empirically found that instead of mean square error, cosine distance is more suitable in the embedding space for distance calculation. The objective function related to this task is
\begin{equation}
    \textnormal{min COS} (G, E) = \frac{1}{m} \sum_{i = 1}^m \left[1 - \frac{E(G(\mathbf{\mathbf{z}}_n^i)) \cdot \mathbf{\mathbf{z}}_n^i}{\lVert E(G(\mathbf{\mathbf{z}}_n^i)) \rVert \lVert \mathbf{\mathbf{z}}_n^i \rVert}\right]
\end{equation}
where $m$ is the mini-batch size.

\paragraph{\textbf{Clustering-specific loss}}

We introduce a clustering-specific loss to learn cluster-friendly representation. For that, we employ cross-entropy (CE) loss, which is computed between $\mathbf{z}_c$ and the soft-max layer output $\hat{\mathbf{z}}_c$ of the encoder network. This loss along with the GAN mini-max objective and the latent variable recovery loss in $\mathbf{z}_n$ encourages clustering in the latent space and also increases discriminative information. We minimize the cross-entropy between the predicted result and the ground truth as 
\begin{equation}
   \textnormal{min CE} (G, E) = \frac{1}{m} \sum_{i = 1}^m \left[p(\mathbf{z}_{c, i}^k) \textnormal{log} \hspace{2pt} p\left(E(G(\mathbf{z}_{c, i}^k))\right)\right] 
\end{equation}
where the first term is the empirical probability that the embedding belongs to the $k$-th speaker, and the second term is the predicted probability that the encoder produced embedding belongs to the $k$-th speaker.

\paragraph{\textbf{Joint training}}

The GAN and the encoder networks training in this approach involves joint parameter updates. The final training objective has the following form:
\begin{equation}
\small
    \underset{G, E}{\textnormal{min}} \hspace{2pt} \underset{D} {\textnormal{max}} \left[w_1 \cdot U_{\textnormal{IWGAN}}(D, G) + w_2 \cdot \textnormal{COS} (G, E) + \\ w_3 \cdot \textnormal{CE} (G, E) \right]
\end{equation}
Weights $w_2$ and $w_3$ represent relative significance of preserving continuous and discrete portions of the latent variable. Algorithm \ref{algo1} lists the whole ClusterGAN training procedure.

\begin{algorithm}[!t]
\caption{ClusterGAN training. Default values: $\lambda$ = 10, $m$ = 128, $n_{\textnormal{critic}}$ = 5, $\alpha$ = $1\mathrm{e}{-4}$, $\beta_1$ = 0.5, $\beta_2$ = 0.9}
\label{algo1}
\begin{algorithmic}[1]
\Require{$\lambda$: gradient penalty coefficient; $\alpha$: learning rate; $m$: batch size; $N_{\textnormal{it}}$: number of iterations; $n_{\textnormal{critic}}$: number of critic iterations for each generator iteration; $\alpha$, $\beta_1$, $\beta_2$: Adam hyper-parameters} {}

\For{$it$ = 1 to $N_{\textnormal{it}}$}
\For{$\tau$ = 1 to $n_{\textnormal{critic}}$}
\State Sample $\{\mathbf{x}^{(i)}\}_{i = 1}^m$, a batch of x-vectors
\State Update the discriminator parameters by
\State\vspace*{-\baselineskip}
          \begin{fleqn}[\dimexpr\leftmargini-\labelsep]
          \setlength\belowdisplayskip{0pt}
        \begin{equation*}
        \small
            \begin{multlined} 
            \hspace{17pt}\theta \leftarrow \textnormal{Adam}[\triangledown_{\theta} \{\frac{1}{m}\sum_{i = 1}^m w_1 \cdot [D_{\theta}(\mathbf{x}^{(i)}) - D_{\theta}(G_{\phi}(\mathbf{z}^{(i)})) \\ + \vspace{-10pt}\lambda \cdot \textnormal{GP}]\}, \theta, \alpha, \beta_1, \beta_2]
            \end{multlined}
        \end{equation*}
        \end{fleqn}
        \normalsize
\EndFor

\State Sample $\{\mathbf{z}^{(i)}\}_{i = 1}^m$, a batch of latent vectors
\State Update the generator and encoder parameters by 
\State\vspace*{-\baselineskip}
          \begin{fleqn}[\dimexpr\leftmargini-\labelsep]
          \setlength\belowdisplayskip{0pt}
        \begin{equation*}
        \small
            \begin{multlined} 
            \hspace{5pt} \phi, \psi \leftarrow \textnormal{Adam}[\triangledown_{\phi, \psi} \{ \frac{1}{m}\sum_{i = 1}^m - w_1 \cdot D_{\theta}(G_{\phi}(\mathbf{z}^{(i)})) +  \\ w_2 \cdot \textnormal{COS} (G_{\phi}, E_{\psi}) + w_3 \cdot \textnormal{CE} (G_{\phi}, E_{\psi})\}, \phi, \psi, \alpha, \beta_1, \beta_2 ] 
            \end{multlined}
        \end{equation*}
        \end{fleqn}
        \normalsize
\EndFor
\end{algorithmic}
\end{algorithm}

\subsubsection{Meta-learning}

Thus far we have discussed the training procedure of ClusterGAN, which is considered as a pre-training part of MCGAN training. In the second phase of MCGAN training, we fine-tune the pre-trained encoder with meta-learning based prototypical loss.

\paragraph{\textbf{Meta-learning using prototypical networks}}

Prototypical networks, or protonets, apply a simpler inductive bias (in the form of class prototypes) as compared to other metric-learning based methods and achieve state-of-the-art few-shot performance in image classification \cite{snell2017prototypical} and natural language processing \cite{yu2018diverse}. The key assumption is that there exists an embedding in which samples from each class cluster around a single prototype representation of that class. Protonets learn a non-linear transformation into an embedding space, where every class is represented by its prototype, sample mean of its support set in the embedding space. During inference, an embedded query sample is assigned to its nearest prototype. The encoder network from ClusterGAN is our protonet. 

\paragraph{\textbf{Motivation for fine tuning}}

The motivation behind fine-tuning the encoder with prototypical loss is that it has good generalization ability at test-time to new classes (unseen during training) given only a handful of examples of each new class \cite{snell2017prototypical}. Similar to this setting, in speaker diarization, a trained model for embedding extraction is asked to do clustering among unseen speakers within an audio stream. This is close to a metric learning task, where input audio must be mapped to a discriminative embedding space. Furthermore, a speaker embedding such as x-vector is trained on a speaker classification loss, 
which is not explicitly designed to optimize embedding similarity. Metric learning related losses such as contrastive loss \cite{chopra2005learning}, triplet loss \cite{schroff2015facenet} can resolve the above issues. Nonetheless, these methods require careful pair or triplet selection, which is sometimes time-consuming and performance-sensitive. In this context, prototypical loss trained for learning a metric space to mimic the test scenario might be handy in capturing information related to both generalization and clustering objectives. 

\paragraph{\textbf{Episode training}}

The encoder or the protonet in the MCGAN is trained episodically, where each episode is one mini-batch consisting of $N_C$ categories randomly sampled from total $K$ categories (here, speakers). The mini-batch also contains a labeled set of examples (\emph{support} set $S$) and unlabeled data (\emph{query} set $Q$) to predict classes. Consider the support set $S$ of $N$ labeled examples as $S = \{\mathbf{x}_i, y_i\}_{i = 1}^N$, where each sample $\mathbf{x}_i$ is a $D$-dimensional x-vector in our case and the corresponding speaker label $y_i \in \{1, \ldots , K\}$. We denote $S_k \subseteq S$ as the set of examples labeled with class $k$. The protonet learns a non-linear mapping $f_{\mathbf{\psi}} : \mathbb{R}^D \rightarrow \mathbb{R}^M$. The $M$-dimensional prototype of each class is computed as the mean of the embedded support points belonging to that class 
\begin{equation}
    \mathbf{p}_k = \frac{1}{|S_k|} \sum_{(\mathbf{x}_i, y_i) \in S_k} f_{\mathbf{\psi}}(\mathbf{x}_i) \label{eq1}
\end{equation}
where $\mathbf{\psi}$ is the learnable parameters of the encoder.
\par
During training, every query sample $\{\mathbf{x}_j, y_j\} \in Q$ is classified against $K$ speakers based on a soft-max over the distances to each speaker prototypes in the new embedding space:
\begin{equation}
    p_{\mathbf{\psi}}(y = y_j|\mathbf{x}_j) = \frac{\textnormal{exp}\big(-d(f_{\mathbf{\psi}}(\mathbf{x}_j), \mathbf{p}_{y_j})\big)}{\sum_{k^\prime}\textnormal{exp}\big(-d(f_{\mathbf{\psi}}(\mathbf{x}_j), \mathbf{p}_{k^\prime})\big)} \label{eq2}
\end{equation}
where $d(.)$ represents a distance function. The choice of $d(.)$ can be arbitrary. However, it is shown in \cite{snell2017prototypical} that the squared Euclidean distance, which is a particular class of distance function known as Bregman divergence, is good for the clustering problem, and the training algorithm is equivalent to modeling the supports using Gaussian mixture density estimation. 
Therefore, we also use Euclidean distance as our distance function for proto-learning in the embedding space. The loss function for each mini-batch is the negative log probability for the true class via gradient descent. The prototypical loss within a mini-batch can be written as 
\begin{equation}
    J_{PTL} = \sum_{\{\mathbf{x}_j, y_j \in Q\}} - \textnormal{log} \hspace{2pt} p(y = y_j|\mathbf{x}_j) \label{eq3}
\end{equation}

\vspace{-8pt}
\paragraph{\textbf{Extension to whole training set}}

Suppose the whole training set $\mathcal{D} = \{(\mathbf{x}_1, y_1), \ldots , (\mathbf{x}_{N_{tr}}, y_{N_{tr}}\}$, where each $y_i \in \{1, \ldots , K\}$. Here, $K$ is the total number of speakers in the training set. We iterate through each episode and in each episode, we randomly sample $N_C$ speakers from total $K$ speakers. For each chosen speaker, $N_S$ number of random samples is selected as the support set and from the rest of the samples of that particular speaker, $N_Q$ number of samples is selected as the query set without replacement. The supports are used to construct the class prototypes using Eq. (\ref{eq1}) and the loss is computed with weight updates based on the query samples according to Eq. (\ref{eq2}), (\ref{eq3}). To increase robustness, instead of using fix total number of speakers, we randomly chose the total number of speakers within an episode. The episodic training procedure is illustrated in Algorithm \ref{algo2}.

\begin{algorithm}[!t]
\caption{Meta-learning training for prototypical networks. $N_{tr}$ = number of labeled examples in the training set, $K$ = total number of speakers in the training set, $N_C \leq K$ is the number of speakers per episode, $N_S$ = number of support examples per chosen speaker, $N_Q$ = number of query examples per chosen speaker. $\mathcal{R}(S, N)$ denotes a set of $N$ elements sampled uniformly at random from set $S$, without replacement.}
\label{algo2}
\begin{algorithmic}[1]
\Require{The whole training set $\mathcal{D} = \bigcup_{k=1}^K \mathcal{D}_k$, where $\mathcal{D}_k$ represents the subset of $\mathcal{D}$ containing all elements such that $\{(\mathbf{x}_i, y_i); y_i = k\}$}
\State $N_C \leftarrow \mathcal{R}(\{10, 20, \ldots , 150\}, 1)$
\algorithmiccomment{Randomly select total number of speakers in an episode}
\State $V \leftarrow \mathcal{R}(\{1, \ldots , K\}, N_C)$ \algorithmiccomment{Randomly select speakers in an episode}
\For{$k$ in $\{1, \ldots , N_C\}$} 
\State $S_k \leftarrow \mathcal{R}(\mathcal{D}_{V_k}, N_S)$ \algorithmiccomment{Supports}
\State $Q_k \leftarrow \mathcal{R}(\mathcal{D}_{V_k} \backslash S_k, N_Q)$ \algorithmiccomment{Queries}
\State $\mathbf{p}_k \leftarrow \frac{1}{N_S} \sum_{(\mathbf{x}_i, y_i) \in S_k} f_{\mathbf{\psi}}(\mathbf{x}_i)$ \algorithmiccomment{Prototypes}
\EndFor
\State $J_{PTL} \leftarrow 0$
\For{$k$ in $\{1, \ldots , N_C\}$}
\For{$(\mathbf{x}_j, y_j)$ in $Q_k$}
\State\vspace*{-\baselineskip}
          \begin{fleqn}[\dimexpr\leftmargini-\labelsep]
          \setlength\belowdisplayskip{0pt}
        \begin{equation*}
        \small
            \begin{multlined} 
            \small
            \hspace{15pt} J_{PTL} \leftarrow J_{PTL} + \frac{1}{N_C N_Q} [-d(f_{\mathbf{\psi}}(\mathbf{x}_j), \mathbf{p}_{y_j}) + \\ \hspace{10pt} \mathrm{log} \sum_{k^\prime} \mathrm{exp}\big(-d(f_{\mathbf{\psi}}(\mathbf{x}_j), \mathbf{p}_{k^\prime})\big)]  \hspace{20pt} $\algorithmiccomment{\normalsize Loss update}$
            \end{multlined}
        \end{equation*}
        \end{fleqn}
        \normalsize
\EndFor
\EndFor
\end{algorithmic}
\end{algorithm}

\vspace{-15pt}
\subsection{MCGAN testing}

After completion of offline training, only the trained encoder model in MCGAN is used to produce the proposed latent embeddings for the input x-vectors of a given test diarization session. The concatenated latent embeddings ($\mathbf{z}_n$ and $\mathbf{z}_c$) for ClusterGAN or logits for MCGAN are clustered using k-means or NME-SC, and speaker labels of each audio segment are obtained.

\vspace{-15pt}
\subsection{Normalized Maximum Eigengap Spectral Clustering (NME-SC)}

We adopt NME-SC\footnote{\url{https://github.com/tango4j/Python-Speaker-Diarization}} as our spectral clustering method in this paper for speaker diarization evaluation in the unknown number of speakers condition. The NME-SC algorithm can auto-tune parameters of the clustering and also provides improved speaker diarization performance as compared to traditional spectral clustering approaches. As reported in \cite{park2019auto}, the steps to perform NME-SC are: (a) Construct affinity matrix ($\mathbf{A}$) based on cosine similarity values between the segment embeddings. (b) Binarize $\mathbf{A}$ based on a $p$-value by converting the $p$-largest elements in each row of $\mathbf{A}$ to 1 and else to 0. 
(c) Perform symmetrization on the binarized affinity matrix $\mathbf{A}_p$ to obtain $\mathbf{\bar{A}}_p$ and compute the Laplacian matrix $\mathbf{L}_p$ as $\mathbf{L}_p = \mathbf{D}_p - \mathbf{\bar{A}}_p$, where $\mathbf{D}_p$ is a diagonal matrix and $\mathbf{D}_p = \sum_{j = 1}^n \mathbf{\bar{A}}_{p, ij}$
(d) Perform eigen decomposition on $\mathbf{L}_p$ 
and create eigengap vector ($\mathbf{e}_p)$. (e) Perform NME analysis to 
estimate the optimum value $\hat{p}$ and number of clusters $k$ for a given session.
(f) Select the $k$-smallest eigenvalues and the corresponding eigenvectors to construct a matrix $\mathbf{P} \in \mathbb{R}^{n \times k}$. (g) Cluster the row vectors of $\mathbf{P}$ using k-means algorithm. The details of the NME-SC algorithm is precisely described in \cite{park2019auto}. 


\vspace{-12pt}
\section{Database description}\label{section4}

We evaluate our proposed speaker diarization system on five different diverse databases covering many possible types and domains that are encountered in real-world scenarios.

\vspace{-12pt}
\subsection{CALLHOME database}

CALLHOME contains telephonic conversations with sampling frequency 8 kHz. In speaker diarization literature, NIST 2000 speaker recognition evaluation challenge disk-8 is referred to as CALLHOME \cite{martin2001speaker}. It is a multi-lingual database distributed across six languages: English, Spanish, Arabic, Mandarin, Japanese, and German. The database comprises 500 conversations with the number of speakers in each session varying from 2 to 7. The telephone recordings are of 1 to 10 min duration and their distribution of the number of speakers is given in \cite{shum2013unsupervised, senoussaoui2014study}.

\vspace{-12pt}
\subsection{AMI database}

The AMI is a publicly available meeting corpus consists of 171 recordings and 100 hours of data\footnote{\url{http://groups.inf.ed.ac.uk/ami/download/}}. The meetings are recorded at four different sites (Edinburgh, Idiap, TNO, and Brno). We use the multiple close-talk microphone data post beamforming. For our evaluation, we follow the official speech recognition partition of AMI database with TNO meetings excluded from dev and eval set. The same split is also used in \cite{sun2019speaker, li2019discriminative}. The train, dev, and eval splits have no speaker overlap and the details are shown in Table \ref{table1}.

\vspace{-10pt}
\subsection{DIHARD II database}

The DIHARD II database is from the DIHARD challenge conducted in 2019. It is a multi-domain database focusing on difficult speaker diarization settings. The database is comprised of diverse recordings collected from domains like meeting speech, restaurant recordings, child language acquisition recordings, YouTube videos, clinical recordings, etc \cite{ryant2019second}. The DIHARD challenge features two audio input conditions:  single-channel and multi-channel. We evaluated our system on single-channel data with reference SAD, which is track 1 in the challenge. Moreover, the database has two subsets: development and evaluation. In this work, speaker diarization performance of proposed and other baseline systems are compared only on the development part of the database. The development set contains 192 recordings typically are of short duration ($<$ 10 min) sessions with the number of speakers in each session varies from 1 to 10.

\begin{table}[!t]
\vspace{-10pt}
\caption{Details of the AMI data set used for our experiments.}
\vspace{-8pt}
\centering
\setlength{\tabcolsep}{10pt}
\begin{tabular}{c|cc}
\Xhline{2.5\arrayrulewidth}
      & \#Meetings & \#Speakers          \\ \Xhline{2.5\arrayrulewidth}
Train & 136     & 155                 \\
Dev   & 14      & 17  \\
Eval  & 12      & 12  \\ \Xhline{2.5\arrayrulewidth}
\end{tabular}\label{table1}
\vspace{-18pt}
\end{table}

\vspace{-12pt}
\subsection{ADOS and BOSCC databases}

The proposed system is also tested on two special child-clinician interaction corpora 
from the sample of children with autism spectrum disorder (ASD). ADOS (Autism Diagnosis Observation Schedule) is a diagnostic tool based on expert clinical administration and observation that produces a diagnostic algorithm score to inform clinical diagnosis of ASD. ADOS comprises 14 play-based conversational tasks, from within which we select two sub tasks: \emph{Emotion and Social Difficulties and Annoyance} from 272 sessions for our evaluation. Each of these dyadic sessions is of duration $<$ 10 min.  BOSCC (Brief Observation of Social Communication Change) is a behavioral observation based autism treatment outcome measure that includes play-based conversational segments of dyadic interaction between a child and an adult (e.g., examiner or caregiver) \cite{grzadzinski2016measuring}. 
In this work, the diarization performance is tested on 24 BOSCC sessions that were collected in a clinical setting. A BOSCC session typically lasts for 12 minutes. The ADOS and BOSCC data considered here are from verbal children and adolescents with autism.

\vspace{-12pt}
\section{Experimental setup}\label{section5}

\subsection{Speech segmentation}

In all the experiments, we have used uniform segmentation (as followed in Kaldi) on the speech intervals specified by the oracle SAD. All the experiments reported in this paper use oracle SAD, which is also a common practice in speaker diarization research \cite{garcia2017speaker, zhang2019fully, diez2019analysis}. Since our focus is on the effectiveness of proposed embeddings in speaker clustering, we use oracle SAD to eliminate the chance of introducing undesirable error initially due to potential performance uncertainty in automated system SAD. For all the experiments, a sliding window of 1.5 sec duration and overlap of 1 sec is employed to produce speaker-homogeneous segments. Note that in this work no re-segmentation module is applied in the final processing step.

\vspace{-12pt}
\subsection{x-vector extraction}

We use x-vectors from the CALLHOME\footnote{\url{https://kaldi-asr.org/models/m6}} and Voxceleb\footnote{\url{https://kaldi-asr.org/models/m7}} recipe as pre-trained audio embeddings for the 8 kHz and 16 kHz data, respectively. 
The x-vector dimensions are of 128 and 512 for 8 kHz and 16 kHz audio data, respectively.

\vspace{-14pt}
\subsection{ClusterGAN model specifications}

We train two different ClusterGAN models to evaluate diarization performance on various databases. To test speaker diarization performance in CALLHOME which contains 8 kHz telephonic conversations, we train ClusterGAN network in a supervised manner based on AMI train (downsampled to 8 kHz) and switchboard (NIST SRE 2000, disk-6) data. 
The other model which we employ for diarization performance evaluation on all other databases (AMI, DIHARD II dev, ADOS, BOSCC) containing 16 kHz data is trained on AMI train and ICSI data. We use 60 beamformed ICSI \cite{janin2003icsi} sessions with a total number of 46 speakers. 
The architectures details of generator ($G$), discriminator ($D$) and encoder ($E$) networks in ClusterGAN are  shown in Table \ref{archi}. 
Moreover, we set the learning rate to 1e-4 and adopt Adam optimization with a mini-batch size of 128 samples to optimize the three networks. We choose the weights $w_1$, $w_2$, and $w_3$ as 1, 10, and 10, respectively, by tuning the diarization error rate (DER) on a held-out set for the 8 kHz model and AMI dev set for the 16 kHz model. 
It is to be noted that all the above-mentioned model specifications are kept the same for all the experiments reported in this paper. 

\vspace{-14pt}
\subsection{MCGAN specifications}

We fine-tune the prototypical network, i.e., the pre-trained encoder in ClusterGAN using Euclidean distance based prototypical loss, which is found to be more effective than cosine distance in \cite{snell2017prototypical}. We use the same encoder for embedding extraction for both support and query points; while x-vectors from the training data form the support and queries. We fine-tune the pre-trained encoder by freezing its first two hidden layers and then train it with prototypical loss. We develop support and query set from the same training data that are used to train the ClusterGAN. Instead of using the fixed number of classes to construct all the episodes, we randomly choose the number of classes from 10 to 150 with intervals of 10 per training episode and found this approach is slightly more effective. The number of shots to use in the support set is selected by tuning the DER on the AMI dev set. We fix the number of supports and queries to 10 for all the experiments. 

\begin{table}[!t]
\vspace{-10pt}
\caption{ClusterGAN architecture details.}
\vspace{-8pt}
\centering
\setlength{\tabcolsep}{2pt}
\begin{tabular}{ccc}
\Xhline{2.5\arrayrulewidth}
Generator ($G$)   & Discriminator ($D$)    & Encoder ($E$)
\\ \Xhline{2.5\arrayrulewidth}
\begin{tabular}[c]{@{}c@{}}Input: Linear, $\mathbf{z} = (\mathbf{z}_n, \mathbf{z}_c)$ \\ $\in \mathbb{R}^{d_\mathbf{z}}$, $d_n^\ast$ = 90, \\ $d_c^{\dagger}$ = 932 for 8 kHz model \\ and 201 for 16 kHz model \end{tabular} & \begin{tabular}[c]{@{}c@{}}Input: Linear, \\ $\mathbf{x} \in \mathbb{R}^{d_{\mathbf{x}}}$ \end{tabular} & \begin{tabular}[c]{@{}c@{}}Input: Linear, \\ $\mathbf{\hat{x}} \in \mathbb{R}^{d_{\mathbf{x}}}$ \end{tabular}                                      \\ \hline
FC$^{\ddagger}$ 512 ReLU                                                                                                                                                             & FC 512 ReLU                                                                                 & FC 512 ReLU                                                                                                                      \\ \hline
FC 512 ReLU                                                                                                                                                             & FC 512 ReLU                                                                                 & FC 512 ReLU                                                                                                                      \\ \hline
\multirow{4}{*}{\begin{tabular}[c]{@{}c@{}}Output: FC $d_\mathbf{x}$ \\ linear for $\hat{\mathbf{x}}$\end{tabular}}                                                                                & FC 512 ReLU                                                                                 & FC 1024 ReLU                                                                                                                     \\ \cline{2-3} 
                                                                                                                                                                        & Output: FC 1 linear                                                                         & \begin{tabular}[c]{@{}c@{}}Output: FC $d_\mathbf{z}$ linear \\ for $\hat{\mathbf{z}}$. Softmax on \\ last $d_c$ to obtain $\hat{\mathbf{z}}_c$\end{tabular} \\ \Xhline{2.5\arrayrulewidth}
\multicolumn{3}{l}{\footnotesize{$^\ast$Dimension of $\mathbf{z}_n$, $^{\dagger}$Dimension of $\mathbf{z}_c$, $^{\ddagger}$Fully-connected}}\\
\end{tabular}\label{archi}
\vspace{-15pt}
\end{table}

\vspace{-14pt}
\subsection{Baseline systems}

We compare our proposed embeddings with different back-end clustering techniques against several baselines and state-of-the-art diarization systems in five different databases. Since our proposed system incorporates x-vectors as input features, we use Kaldi-based x-vectors with PLDA scoring and AHC clustering as our main baseline system. Furthermore, we show results for x-vector embedding and k-means or spectral clustering (SC) as back-ends, and these are other baseline systems. For a fair comparison, we also report the results of our proposed embeddings with k-means and SC back-ends. We also perform embedding fusion with x-vectors with k-means and SC back-end clustering. Note that for the oracle number of speakers we used fixed tuned $p$-value binarized SC \cite{park2019auto}, whereas for the estimated number of speakers we adopt NME-SC \cite{park2019auto} for all the experiments in this paper. 

\vspace{-10pt}
\subsection{Performance metrics}

We evaluate the proposed speaker diarization system with NIST diarization error rate (DER) \cite{fiscus2006rich}. Following the approach described in \cite{fiscus2006rich}, we use a collar of 0.25 sec for all the databases DER evaluation, except DIHARD II, where zero collar is used according to the challenge criteria \cite{ryant2019second}. Since we employ oracle SAD in this work, all the reported DER is exactly speaker confusion and no missed or false alarm speech. 

\vspace{-10pt}
\section{Experimental results}\label{section6}

\subsection{CALLHOME database}

\subsubsection{Ablation study}

\begin{figure}[!t]
\centering
\includegraphics[width= 0.5\textwidth]{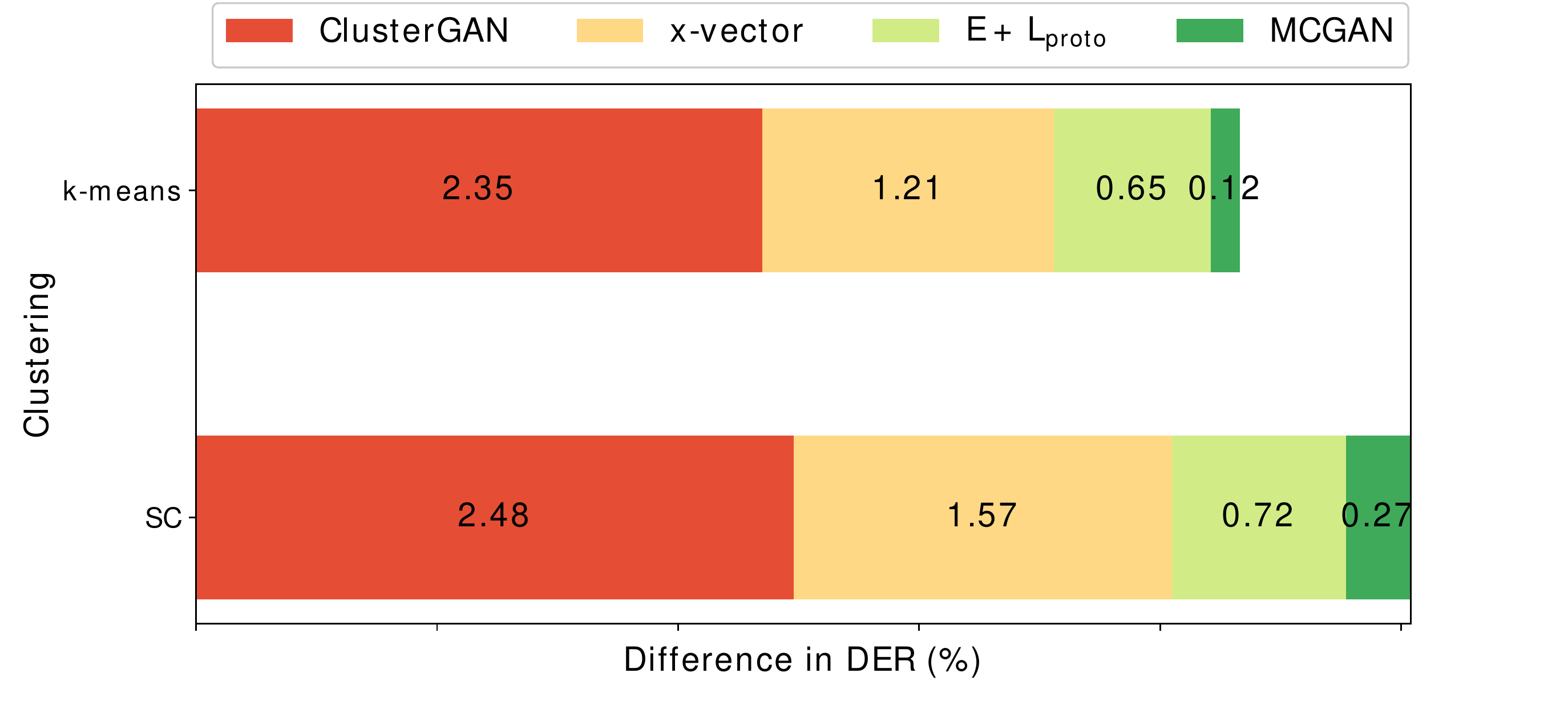}
\vspace{-18pt} 
\caption{Difference in DER (\%) when the final proposed system (fused) is trained using some components of the proposed system.}\label{fig3}
\vspace{-12pt}
\end{figure}

We perform an ablation study to examine the contribution of each component of our proposed system. To do so, we train a single encoder network with random initialization based on prototypical loss only ($E + L_{\textnormal{proto}}$) using x-vectors of the training data as input. We compute DER for different embeddings like ClusterGAN, $E + L_{\textnormal{proto}}$, MCGAN, x-vector, x-vector + MCGAN with both k-means and spectral clustering (SC) as back-end clustering. Fig.~\ref{fig3} shows the difference in DER values between each embedding and our final proposed embedding (x-vector + MCGAN) with two mentioned clustering techniques in CALLHOME. The mean difference for all the sessions is shown between each scenario and the final proposed setting.
\par
It is observed from the figure that all the sub-components contribute to improving DER performance. The effect of the components on diarization performance in the CALLHOME dataset with increasing order is ClusterGAN, x-vector, $E + L_{\textnormal{proto}}$, and MCGAN for both the k-means and SC back-ends. For both k-means and SC back-ends, $E + L_{\textnormal{proto}}$ is more effective than x-vector. Moreover, the figure shows fine tuning ClusterGAN with prototypical loss (MCGAN) is important in achieving improved DER. It also demonstrates that meta training and embedding fusion are the key components to finally obtain the best results in CALLHOME. 

\begin{table}[!t]
\centering
\caption{Results on CALLHOME database for MAPD of predicted speaker number and POC in estimating speaker number.}
\vspace{-8pt}
\centering
\setlength{\tabcolsep}{6pt}
\begin{tabular}{ccccc}
\Xhline{2.5\arrayrulewidth}
Metric    & x-vector & ClusterGAN   & MCGAN & Fusion \\
&&&&(x-vector + MCGAN)\\
\Xhline{2.5\arrayrulewidth}
MAPD       & 12.54\%  & 11.23\% & \textbf{9.76\%}  & 10.59\%                            \\
POC     & 74.15\% & 72.14\%  & \textbf{75.55\%}   & 75.35\%                          \\

\Xhline{2.5\arrayrulewidth}
\end{tabular}\label{table2}
\vspace{-12pt}
\end{table}

\subsubsection{Number of predicted speakers}

In addition to DER, the mean absolute percentage deviation (MAPD) of the predicted number of speakers and percentage of the correct number of speaker estimation (POC) across all the sessions are also useful metrics in the context of the number of speakers estimation in speaker diarization. The lower the MAPD and the higher the POC is the better speaker estimation. The results on CALLHOME are summarized in Table \ref{table2}. It is evident from the table that MCGAN embeddings are more robust and accurate in estimating the number of speakers than ClusterGAN embeddings and x-vectors. The performance of fused embeddings is slightly worse than MCGAN. Therefore, it is expected that MCGAN will perform better than ClusterGAN and x-vector in the estimated number of speakers condition.

\subsubsection{Overall performance evaluation} \label{overall}

In this section, we present the experimental results on the whole CALLHOME evaluation set by using the tuned parameters of the different versions of our proposed diarization system. We compare the proposed system with other baselines and recent state-of-the-art diarization methods. The experimental results for both known and unknown numbers of speakers are reported in Table \ref{table3}. 
Note that for known or oracle number of speakers we use fix $p$-value which is tuned on Kaldi CALLHOME-1 held-out set and apply it to CALLHOME-2 and vice versa.

\begin{table}[!t]
\centering
\scriptsize
\caption{Results on CALLHOME database for the baseline and proposed systems.}
\vspace{-8pt}
\centering
\setlength{\tabcolsep}{0.5pt}
\begin{tabular}{cccc}
\Xhline{2.5\arrayrulewidth}
Embedding    & Back-end    & \setlength{\tabcolsep}{-2pt} \begin{tabular}[c]{@{}c@{}}Avg. DER (\%) \\ (oracle SAD, \\ known \#speakers)\end{tabular} & \setlength{\tabcolsep}{-5pt} \begin{tabular}[c]{@{}c@{}}Avg. DER (\%) \\ (oracle SAD, \\ est. \#speakers)\end{tabular} \\
\Xhline{2.5\arrayrulewidth}
x-vector       & \multirow{6}{*}{k-means} & 9.00                         & 8.69                            \\
ClusterGAN     &        & 10.24                         & 9.83                           \\
$E + L_{\textnormal{proto}}$ &  & 9.09                          & 8.13                           \\
MCGAN          &   & 8.72                         & 7.60                            \\
x-vector + ClusterGAN &  & 8.98                          & 8.77                           \\
x-vector + MCGAN &  & 8.40                          & 7.48                           \\ \hline
x-vector       & \multirow{6}{*}{SC}      & 6.23                         & 8.32                            \\
ClusterGAN     &       & 7.62                         & 9.24                           \\
$E + L_{\textnormal{proto}}$ &  & 6.34                          & 7.48                           \\
MCGAN     &        & 6.01                          & 7.03                            \\
x-vector + ClusterGAN &  & 6.22                          & 7.70                           \\
x-vector + MCGAN &  & \textbf{5.73}                          & \textbf{6.76}                           \\
\Xhline{2.5\arrayrulewidth}
Wang et al. \cite{wang2018speaker} d-vector & SC & -- & 12.00 \\
Romero et al. \cite{garcia2017speaker} x-vector & PLDA+AHC+VB$^\ast$ & -- & 9.90 \\
Kaldi x-vector      & PLDA+AHC+CV$^\dagger$ & 7.12                         & 8.39                            \\
Zhang et al. \cite{zhang2019fully} d-vector (5-fold) & UIS-RNN+CV & -- & 7.60 \\
Park et al. \cite{park2019auto} Kaldi x-vector & NME-SC & -- & 7.29 \\
\Xhline{2.5\arrayrulewidth}
\multicolumn{4}{l}{\footnotesize{$^\ast$Variational Bayes re-segmentation, $^\dagger$Cross-validation}}\\
\end{tabular}\label{table3}
\vspace{-15pt}
\end{table}

\par

\begin{figure*}[!t]
\centering
    \includegraphics[width= \textwidth]{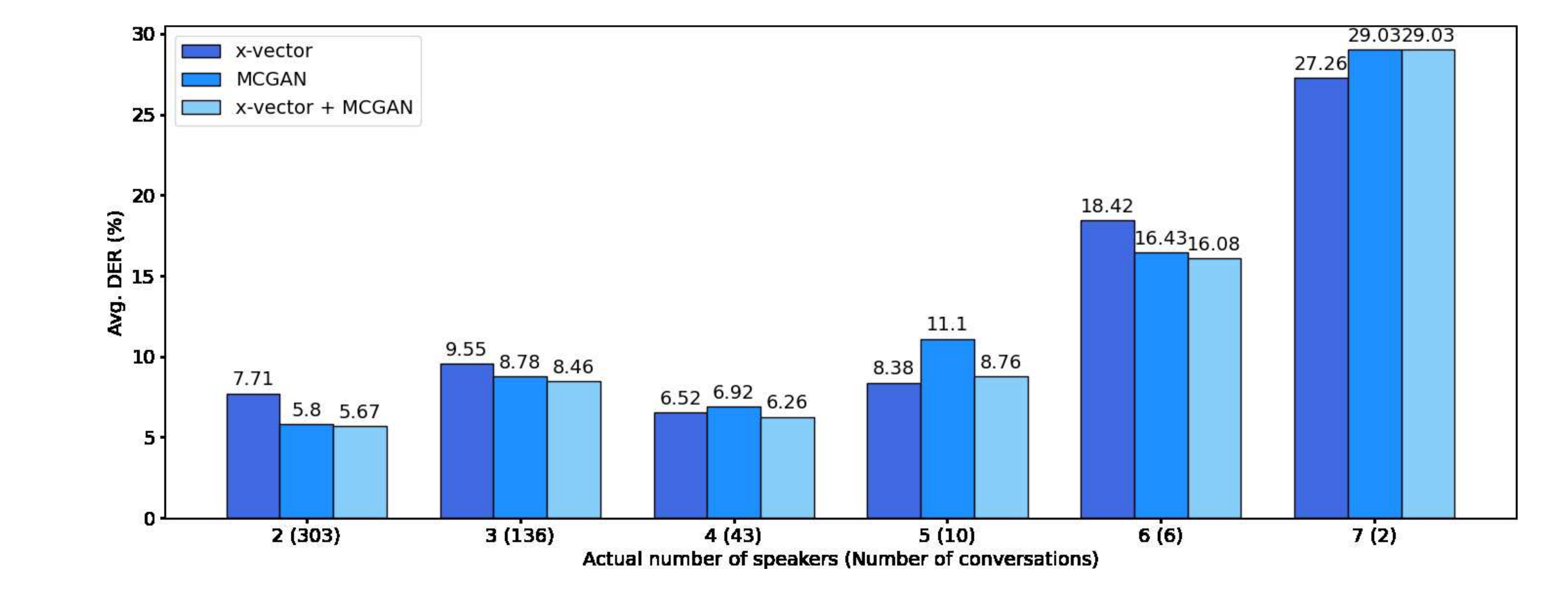}
    \vspace{-30pt} 
    \caption{Avg. DER (\%) analysis with respect to number of speakers in a diarization session.}\label{fig4}
\vspace{-15pt}
\end{figure*}

From Table~\ref{table3} column 3, we observe that for known number of speakers, the ClusterGAN embedding does not outperform x-vectors for both k-means and SC. 
However, we see that MCGAN embeddings which are extracted after fine-tuning the pre-trained encoder with prototypical loss provide superior performance over x-vectors for both k-means and SC back-ends. MCGAN reduces average DER of ClusterGAN from 10.24\% to 8.72\% and from 7.62\% to 6.01\% for k-means and SC, respectively. Therefore, fine-tuning the protonet ($E$) with meta-learning related prototypical loss is useful for better generalization. To corroborate this fact, the average DER of an encoder initialized with random weights and trained with prototypical loss ($E + L_{\textnormal{proto}}$) on the training data is also calculated and the corresponding results are given in Table~\ref{table3} row 3 and 9. It is observed from the table that proposed MCGAN embeddings reduce DER of $E + L_{\textnormal{proto}}$ from 9.09\% to 8.72\% and from 6.34\% to 6.01\% respectively for k-means and SC. Finally, we obtain further improvement in DER by incorporating embedding fusion between x-vector and MCGAN embeddings. We achieve the best DER of 5.73\% for the known number of speakers and SC back-end, which is significantly better than the Kaldi x-vector state-of-the-art (average DER 7.12\%) and also superior to the x-vector with SC (average DER 6.23\%). The relative improvement of our final proposed system over Kaldi state-of-the-art is 19.52\% for known number of speakers. 
\par
We show the diarization performance of all the systems for the estimated number of speakers in Table~\ref{table3} column 4. The number of speakers for k-means and SC is estimated using NME-SC. 
From Table~\ref{table3} column 4, we see a similar trend in performance for estimated number of speakers. The biggest improvement in DER for proposed embeddings comes from MCGAN, embedding fusion, and most importantly the SC. For the same back-end setting, the proposed MCGAN and fused embeddings are better than x-vector. It is important to note that surprisingly in many of the settings (except SC) we obtain reduced DER for the automatically estimated number of speakers case than for oracle number of speakers. This could be attributed to the fact that even though the number of clusters may be correct for the oracle case there might be inherent speaker confusions, whereas, for the estimated number of speakers, the clusters based on data-driven estimation may be purer even if the estimated number of clusters is not exactly correct. The embedding fusion between x-vector and MCGAN with SC back-end yields the best DER value of 6.76\% for the estimated number of speakers with a relative improvement of 19.43\% over the Kaldi x-vector system. The next best system--MCGAN with SC--produces a DER of 7.03\%, which is also significantly better than the Kaldi x-vector and x-vector with SC back-end. 
We also present the recent best system's results that are reported in the literature on the CALLHOME evaluation set. Many of these systems use cross-validation to train or adapt their systems. However, without using any cross-validation, the proposed system outperforms all the recent diarization systems in CALLHOME. 

\begin{figure}[!t]
\centering
    \includegraphics[width= 0.5\textwidth]{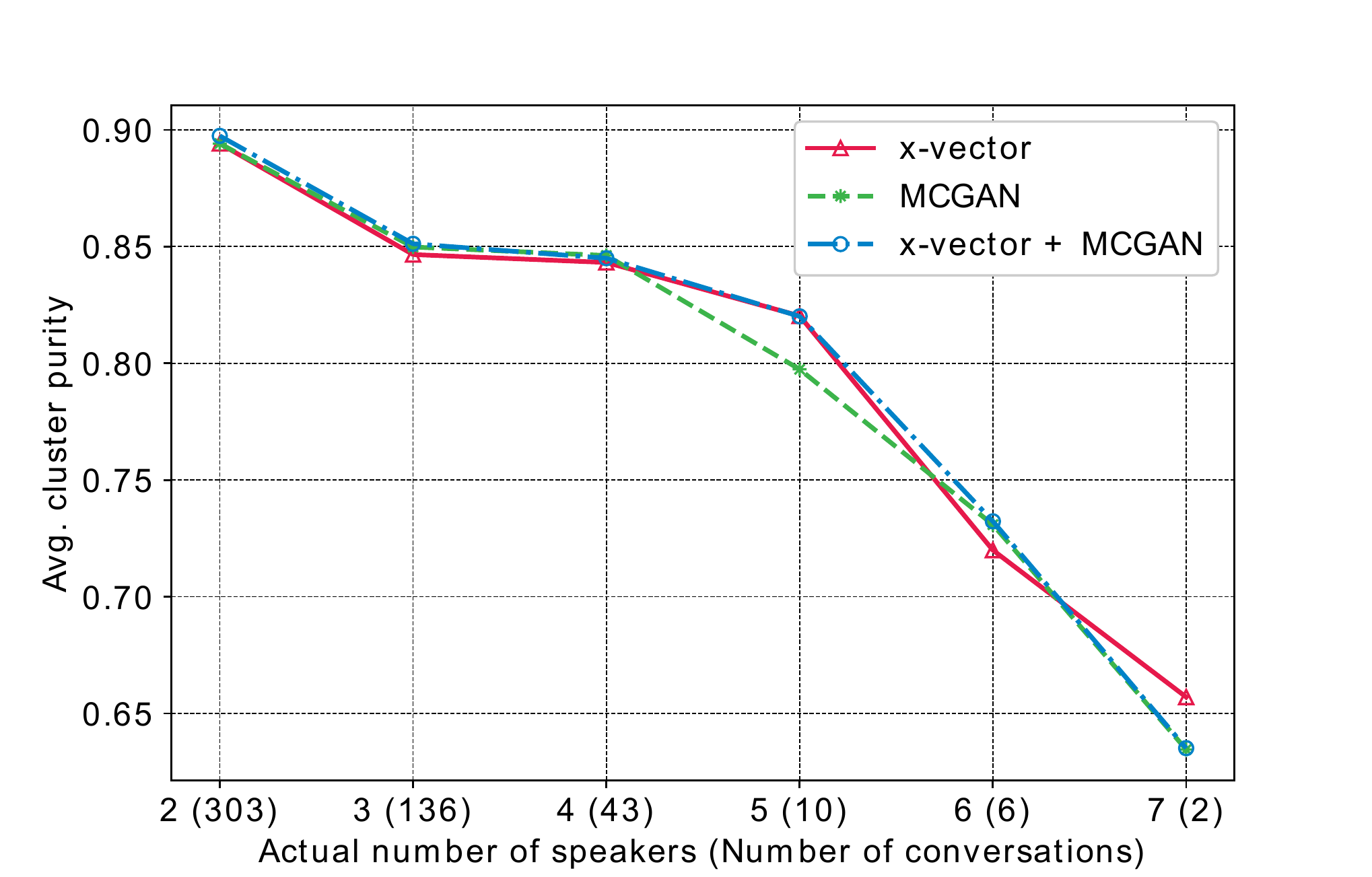}
    \vspace{-15pt} 
    \caption{Average cluster purity analysis with respect to number of speakers in a diarization session.}\label{fig5}
\vspace{-15pt}
\end{figure}

\subsubsection{Analysis of Experimental Results}

We first break down the average DER on CALLHOME database according to the number of speakers. The corresponding DERs are plotted in a group bar plot in Fig.~\ref{fig4} for x-vector, MCGAN and fused (x-vector + MCGAN) embeddings with NME-SC back-end and estimated number of speakers. It is evident from the figure that our proposed MCGAN and fused embeddings achieve significantly better DER values than x-vector for two and three speaker cases. For four and five speakers, x-vector is better than MCGAN. However, the fused embeddings provide better performance than x-vectors for most of the speaker conditions (two, three, four, and six) and this covers majority of the conversations in the dataset. 
In the seven-speaker condition, the fused system is not able to outperform x-vector, which is possibly an anomaly since the number of sessions containing seven speakers is only two in this database. The reason behind obtaining better results using fused embeddings with SC is attributed to the complementary merits of the x-vector and MCGAN embeddings, and the modeling power of the NME-SC algorithm on embeddings. 
\par 
In light of this, it would be reasonable to consider another performance metric related to the clustering mechanism. Fig.~\ref{fig5} shows the average cluster purity with respect to changes in the number of speakers in CALLHOME for x-vector, MCGAN, and fused (x-vector + MCGAN) embeddings with NME-SC back-end and estimated number of speakers. We can see that the fused embeddings provide better cluster purity than x-vector and MCGAN for all the number of speaker cases except 7, which has only two sessions. We observe noticeable improvement for MCGAN and fused over x-vector for the six-speaker case and little improvement in two and three speaker conditions, a result consistent with our DER analysis with respect to the number of speakers.
\par
\begin{figure}[!t]
\centering
    \includegraphics[width= 0.5\textwidth]{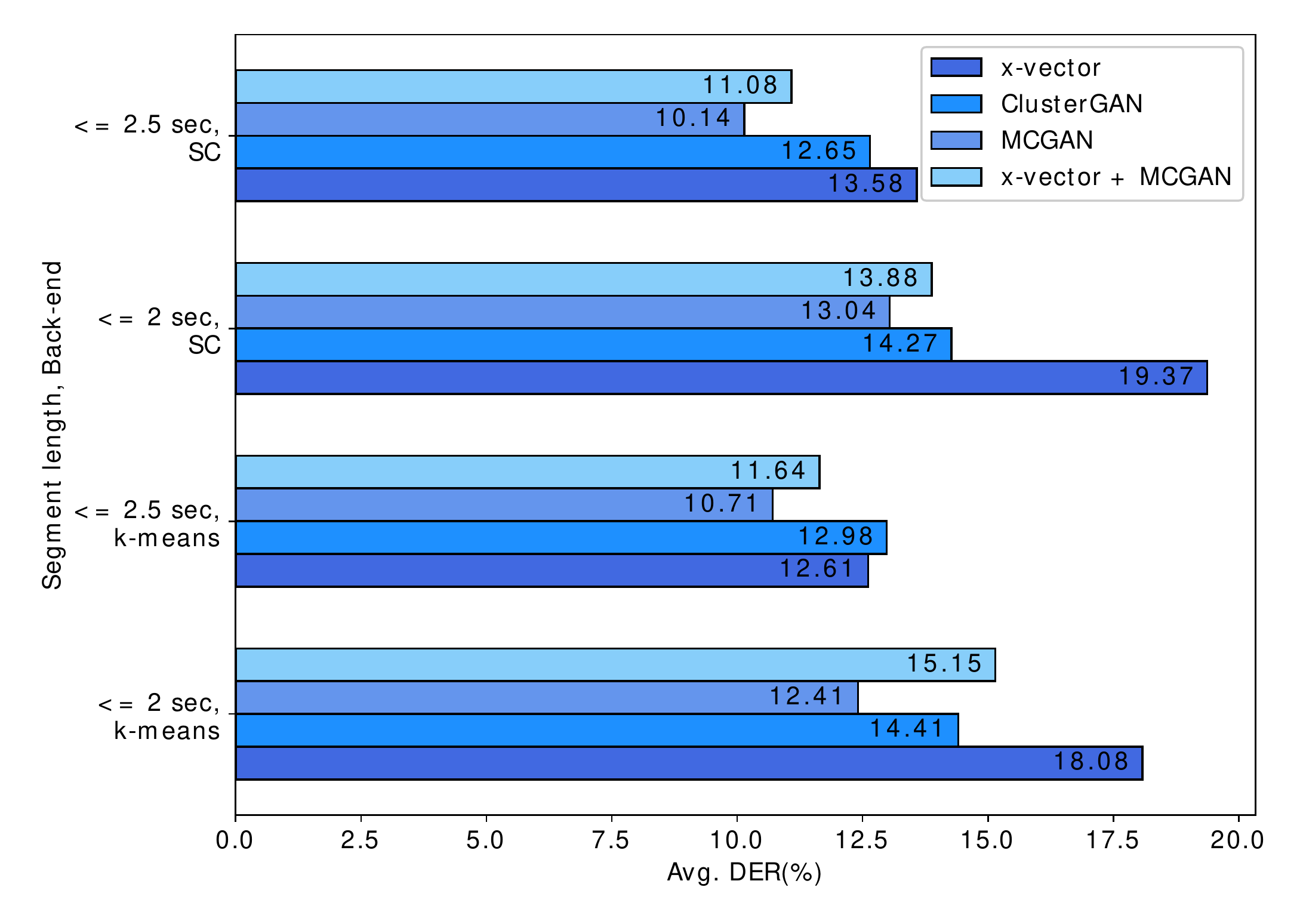}
    \vspace{-22pt} 
\caption{Average DER (\%) analysis on short speech segment diarization sessions.}\label{fig6}
    \vspace{-17pt}
\end{figure}

We extend the analysis by checking the effectiveness of our proposed system in a more challenging practical scenario namely diarization in short speech segment case. Shorter segments usually provide low-quality speaker embeddings. To carry out this analysis, we chose conversations from the CALLHOME evaluation set that have a majority number of short duration ($<=$ 2 sec and $<=$ 2.5 sec) speech segments. Here, we select sessions that have more than 80\% of short speech segments in the entire session. We find that number of such sessions is 58 ($<=$ 2 sec) and 129 ($<=$ 2.5 sec), respectively. We compute and plotted the mean DER of the selected sessions in Fig.~\ref{fig6} for x-vector, ClusterGAN, MCGAN, and fused (x-vector + MCGAN) embeddings with k-means and spectral clustering and estimated number of speakers. It is clear from the figure that among the four embeddings, MCGAN embedding produces the lowest average DER for short speech segment sessions compared to x-vector, ClusterGAN, and fused embeddings, and for both the clustering techniques. The fused and ClusterGAN embeddings yield better performance than x-vector for most of the cases. We obtain worse DER values for $<=$ 2 sec segments than $<=$ 2.5 segments, which is not surprising. Finally, we can conclude that MCGAN embedding is more robust than the other embeddings in short speech segment scenarios. 

\vspace{-10pt}
\subsection{AMI database}

\subsubsection{Performance comparison}

Herein we evaluate the diarization performance of several proposed systems and compare them with Kaldi state-of-the-art and other alternative baselines on the AMI meetings corpus. We use the model trained on AMI train and ICSI data to generate our proposed speaker embeddings.
\par
Results presented in Table~\ref{table4} show speaker diarization performance on both the AMI dev and eval sets for oracle SAD with both known and estimated number of speakers. The hyper-parameters for ClusterGAN and MCGAN models and the $p$-value for spectral clustering are tuned on the dev set and applied on the dev and eval set. From Table~\ref{table4} column 3, we see that for the known number of speakers, all the proposed embeddings (ClusterGAN, MCGAN, x-vector + ClusterGAN, x-vector + MCGAN) significantly outperform x-vector for both k-means and SC back-ends. This is true for both AMI dev and eval sets. Moreover, for various proposed settings, the embedding fusion reduces the DER over individual embeddings and SC further lowers the DER as compared to the k-means counterpart. The fused x-vector + ClusterGAN with SC achieves the best DER value of 3.60\% on AMI eval set for the known number of speakers case. This indicates that the proposed embeddings have complementary information to x-vector embeddings. In addition, the fused embeddings with SC outperform Kaldi x-vector baseline system by a significant margin. 

\par
We show results for all the systems on AMI dev and eval sets for the estimated number of speakers in Table~\ref{table4} column 4. The number of speakers for both k-means and SC is estimated by using the NME-SC algorithm. For the Kaldi x-vector baseline, the number of speakers in a session is estimated based on thresholding on the PLDA scores \cite{garcia2017speaker}. 
It is clear from Table~\ref{table4} column 4 that similar to the oracle case, the proposed embeddings are superior to x-vector in dev and eval set under k-means clustering. MCGAN yields better performance compared to the ClusterGAN. As seen previously, MCGAN embeddings are more robust in the estimated number of speakers condition than ClusterGAN. 
Moreover, as expected, the fused embeddings further improve diarization performance for the k-means back-end. On the other hand, spectral clustering boosts diarization performance further for all the embeddings. This ensures the effectiveness of NME-SC over k-means. 
Besides, the embedding fusion provides further reduction in DER for SC back-end. 
Finally, using our proposed fused system, we achieve significantly better performance on AMI dev and eval set with absolute DER of 5.02\% and 2.87\%, respectively, outperforming Kaldi x-vector baseline diarization system.

\begin{table}[!t]
\caption{Results on AMI dev and eval sets for the baseline and proposed systems.}
\vspace{-8pt}
\centering
\footnotesize
\setlength{\tabcolsep}{3pt}
\begin{tabular}{cccccc}
\Xhline{2.5\arrayrulewidth}
\multirow{2}{*}{Embedding} & \multirow{2}{*}{Back-end} & \multicolumn{2}{c}{\begin{tabular}[c]{@{}c@{}}Avg. DER (\%) \\ (oracle SAD, \\ known \#speakers)\end{tabular}} & \multicolumn{2}{c}{\begin{tabular}[c]{@{}c@{}}Avg. DER (\%) \\ (oracle SAD, \\ est. \#speakers)\end{tabular}} \\
                        \cline{3-4} \cline{5-6} & & Dev                              & Eval                             & Dev                                & Eval                               \\ \Xhline{2.5\arrayrulewidth}

x-vector & \multirow{5}{*}{k-means}                 & 11.94                            & 11.45                            & 12.64                                     & 12.26                                    \\
ClusterGAN &  & 7.64                           & 7.69                            &   11.34                                 &    11.51                                \\
MCGAN  &                       & 5.84                             & 6.14                            &   7.09                                 &      6.09                              \\
x-vector + ClusterGAN &            & 6.62                             & 6.46                            & 9.57                                   & 8.63                                   \\
x-vector + MCGAN &            & 5.64                            & 5.48                             & 6.47                                   &  8.76                                  \\ \hline
x-vector       & \multirow{5}{*}{SC}      & 7.32          & 6.88  & 6.42               & 6.23                            \\
ClusterGAN     &        & \textbf{3.86}         & 3.91 & 6.41                & 8.16                           \\
MCGAN     &        & 5.72              & 4.49 & 5.10            & 5.38                            \\
x-vector + ClusterGAN       &       & 3.93          & \textbf{3.60}  & 6.21              & \textbf{2.87}                            \\
x-vector + MCGAN     &        & 5.49         & 4.23 & \textbf{5.02}                & 4.92                           \\
\Xhline{2.5\arrayrulewidth}
Kaldi x-vector    & PLDA+AHC            & 11.65                            & 11.34                            & 11.08                              & 10.37                              \\ 
\Xhline{2.5\arrayrulewidth}
\end{tabular}\label{table4}
\vspace{-12pt}
\end{table}

\begin{figure}[!t]
\centering
\includegraphics[width= 0.5\textwidth]{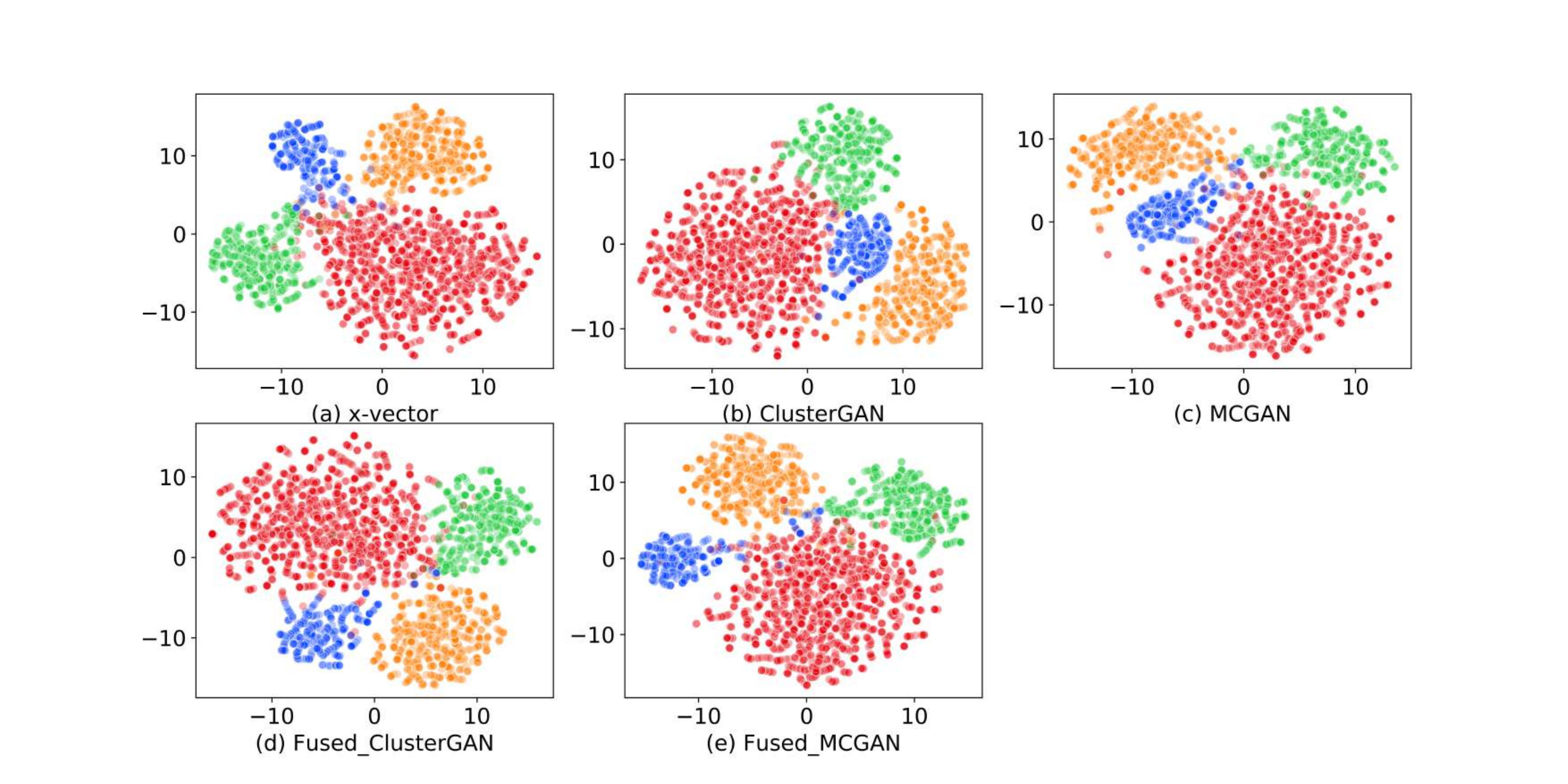}
    \vspace{-20pt} 
    \caption{ t-SNE visualization of (a) x-vector, (b) ClusterGAN, (c) MCGAN (d) x-vector + ClusterGAN, and (e) x-vector + MCGAN embeddings of IS1008a AMI session. This AMI session contains four speakers and each speaker is represented by different colours in the figure.}\label{fig7}
    \vspace{-14pt}
\end{figure}

\subsubsection{t-SNE visualization}

Fig.~\ref{fig7} shows the t-SNE visualization of x-vector, ClusterGAN, MCGAN, x-vector + ClusterGAN and x-vector + MCGAN embeddings. While the proposed ClusterGAN and MCGAN embeddings exhibit compact clusters than x-vectors, MCGAN shows better separation among the clusters than ClusterGAN. It is also evident from the figure that fused embeddings exhibit compact and well-separated clusters. Moreover, in this particular example, x-vector + MCGAN shows near-perfect clustering among the four classes, which is slightly better than x-vector + ClusterGAN.

\vspace{-12pt}
\subsection{DIHARD II database}

\subsubsection{Performance comparisons}

To investigate the effectiveness of the proposed embeddings for speaker diarization in challenging multi-domain settings, we evaluated and report the average DER values on the DIHARD II development database in Table~\ref{table5}. We also compare different versions of proposed embeddings against x-vector with different clustering techniques and existing state-of-the-art systems. 
It is to be noted that the proposed embeddings are based on the same model that was used for AMI database evaluation. This is to check the robustness of our embeddings in real-world noisy scenarios without training explicitly using separate noisy data or data augmentation. Note that for the known number of speakers since we used fixed $p$-value for SC in this paper, we tune it on the same dev set for all the embeddings. It is seen from Table~\ref{table5} column 3 that x-vector produces the best DER value as compared to ClusterGAN and MCGAN for k-means clustering. This is reasonable since the x-vector pre-trained model is based on Voxceleb data with data augmentation. However, embedding fusion with x-vectors improves the performance further for both the proposed embeddings. In contrast to k-means clustering, spectral clustering provides a larger boost in performance for all the embeddings. We attain lower DER value for ClusterGAN than x-vector and MCGAN for known number of speakers and SC back-end. 
\par
We report the DER values for the estimated number of speakers using NME-SC in Table~\ref{table5} column 4. While the performance is not good for k-means with oracle number of speakers, we obtain surprisingly better DER values for the estimated number of speakers. It is seen from the table that MCGAN is superior to ClusterGAN in the estimated number of speakers scenario. However, individually x-vector is better than ClusterGAN and MCGAN in this database. Nonetheless, both the fused embeddings outperform x-vector for k-means back-end. With SC back-end, we achieve significant improvement in performance for all the embeddings. We attain the best DER value of 17.75\% by using x-vector + ClusterGAN embedding and SC back-end. For comparison, we also report the performance of several existing DIHARD challenge II submissions. The challenge top system by BUT achieves a DER value of 18.09\% on the DIHARD II dev set \cite{landini2019but}. However, it is mentioned in the paper that in their system, PLDA was adapted on the same development set. Therefore, it is worthwhile to note that without using PLDA and several additional processing steps like speech enhancement, overlap detection, and variational Bayes (VB) re-segmentation our proposed system outperforms the top system in the DIHARD II development set.
Furthermore, the proposed system is significantly better than the Kaldi x-vector diarization system, which was the baseline in the challenge. 
Thus, the proposed embeddings although extracted from the model trained on AMI train and ICSI data, are promising in terms of generalization, have complementary information to x-vectors, and can yield state-of-the-art performance on a challenging multi-domain database in an embedding fusion set up with the spectral clustering back-end.

\begin{table}[!t]
\caption{Results on DIHARD II development set for the baseline and proposed systems.}
\vspace{-8pt}
\scriptsize
\centering
\setlength{\tabcolsep}{2pt}
\begin{tabular}{cccc}
\Xhline{2.5\arrayrulewidth}
Embedding    & Back-end    & \begin{tabular}[c]{@{}c@{}}Avg. DER (\%) \\ (oracle SAD, \\ known  \#speakers)\end{tabular} & \begin{tabular}[c]{@{}c@{}}Avg. DER (\%) \\ (oracle SAD, \\ est. \#speakers)\end{tabular} \\
\Xhline{2.5\arrayrulewidth}
x-vector       & \multirow{5}{*}{k-means} & 27.12                         & 23.38                            \\
ClusterGAN     &       & 29.68                         & 26.03                           \\
MCGAN          &   & 28.98                         & 25.71                            \\
x-vector + ClusterGAN &  & 27.76                        & 22.87                           \\
x-vector + MCGAN &  & 26.59                          & 22.53      \\  \hline 
x-vector       & \multirow{5}{*}{SC}      & 19.86                         & 18.88                            \\
ClusterGAN     &        & 19.08                        & 21.84                           \\
MCGAN     &        & 19.87                          & 21.16                            \\
x-vector + ClusterGAN &  & \textbf{17.69}                        & \textbf{17.75}                           \\
x-vector + MCGAN &  & 18.86                        & 18.59                           \\
\Xhline{2.5\arrayrulewidth}
Kaldi x-vector      & PLDA+AHC & --                 & 25.59                            \\
Kaldi x-vector      & PLDA(adapted)+AHC & --            & 23.82                            \\
\cite{park2019second} Kaldi x-vector(fusion) & SC & --   & 21.82 \\
\cite{lin2020dihard} Resnet embed. & SC+VB$^\ast$+OD$^\dagger$ & --   & 18.84 \\
\cite{landini2019but} x-vector & PLDA+AHC+VB+OD & --   & 18.09 \\
\Xhline{2.5\arrayrulewidth}
\multicolumn{4}{l}{\footnotesize{$^\ast$Variational Bayes re-segmentation, $^\dagger$Overlap detection}}\\
\end{tabular}\label{table5}
\vspace{-12pt}
\end{table}

\begin{table*}[!t]
\caption{Diarization performance of the proposed and baseline systems in each specific domain on the DIHARD II dev set.}
\vspace{-8pt}
\centering
\setlength{\tabcolsep}{5pt}
\begin{tabular}{cccccccccccc}
\Xhline{2.5\arrayrulewidth}
\multicolumn{1}{c}{System} & \multicolumn{1}{c}{\begin{tabular}[c]{@{}c@{}}audiobooks\\ $^\ast$(12)\end{tabular}} & \multicolumn{1}{c}{\begin{tabular}[c]{@{}c@{}}broadcast\\ \_interview\\ (12)\end{tabular}} & \multicolumn{1}{c}{\begin{tabular}[c]{@{}c@{}}maptask\\ (23)\end{tabular}} & \multicolumn{1}{c}{\begin{tabular}[c]{@{}c@{}}socio\_lab\\ (16)\end{tabular}} & \multicolumn{1}{c}{\begin{tabular}[c]{@{}c@{}}socio\_field\\ (12)\end{tabular}} & \multicolumn{1}{c}{\begin{tabular}[c]{@{}c@{}}court\\ (12)\end{tabular}} & \multicolumn{1}{c}{\begin{tabular}[c]{@{}c@{}}clinical\\ (24)\end{tabular}} & \multicolumn{1}{c}{\begin{tabular}[c]{@{}c@{}}child\\ (23)\end{tabular}} & \multicolumn{1}{c}{\begin{tabular}[c]{@{}c@{}}webvideo\\ (32)\end{tabular}} & \multicolumn{1}{c}{\begin{tabular}[c]{@{}c@{}}meeting\\ (14)\end{tabular}} & \multicolumn{1}{c}{\begin{tabular}[c]{@{}c@{}}restaurant\\ (12)\end{tabular}} \\ \Xhline{2.5\arrayrulewidth}
\begin{tabular}[c]{@{}c@{}}x-vector + \\ NME-SC\end{tabular}                  &   5.92         &    \textbf{4.56}    &    9.23     &  \textbf{7.87}    &   12.17    & 5.60 &    20.20   &    31.65      &    \textbf{33.74}     & 10.20 & 42.68      \\ \hline

\begin{tabular}[c]{@{}c@{}}ClusterGAN + \\ NME-SC\end{tabular}          &    1.98        &    4.75     &   9.50      & 8.75  & 15.54 & 4.67      &   24.48       &   31.77    &     37.37     &  24.94       &    54.04        \\ \hline

\begin{tabular}[c]{@{}c@{}}MCGAN + \\ NME-SC\end{tabular}       &   1.59         &    4.86                                              &  10.14    &   11.99 & 11.64   &    \textbf{4.66}      &   29.33    &    34.26      &    36.38     &   12.50 & 44.41         \\ \hline

\begin{tabular}[c]{@{}c@{}}(x-vector + \\ ClusterGAN) +\\ NME-SC\end{tabular} &   2.29         &    4.68    &  9.10 & 8.20 & \textbf{9.20} &  5.31     & \textbf{16.37}    &  \textbf{29.69}     &   34.66 &  \textbf{9.90} &   \textbf{40.46}         \\ \hline

\begin{tabular}[c]{@{}c@{}}(x-vector + \\ MCGAN) +\\ NME-SC\end{tabular}      &   3.64 &  4.77   &  \textbf{8.92} & 8.63 & 12.28& 4.78&    18.46   &   31.06    &  34.42   &    11.24   &  42.10  \\ \Xhline{2.5\arrayrulewidth}
Lin et al. \cite{lin2020dihard} & \textbf{0.0} & 5.52 & 11.32 &12.08 &14.46 &12.12 & 17.42 &37.38 &34.90 &25.73 &43.00 \\

\Xhline{2.5\arrayrulewidth}
\multicolumn{12}{l}{\footnotesize{$^\ast$ The number of sessions within each domain}}
\end{tabular}\label{table6}
\vspace{-12pt}
\end{table*}

\subsubsection{DER analysis according to the domains}

To understand how our proposed embeddings with spectral clustering behave in each specific domain of DIHARD II dev set, we split the DER according to the context of the database. The results shown in Table~\ref{table6} indicate high variability in performance across the domains. The proposed embeddings (ClusterGAN and MCGAN) individually are not able to outperform  x-vectors except on audiobooks data, which contains only one speaker. 
However, the fused embeddings offer promising performance on most of the domains compared to the x-vectors. The worst performing domains for our embeddings are restaurant, webvideo, and child. The metadata analysis of DIHARD II dev set in \cite{lin2020dihard} shows that restaurant sessions are highly noisy and also contain a large number of speakers. In addition, restaurant and webvideo data comprise significant speech overlap. On the other hand, although the child data contains less amount of overlap, the observed worse performance is because the children in the sessions are 6-18 months old, and have high variability in their speech; moreover, more than two speakers are present in those sessions. We also compare our results with a recently proposed diarization system on DIHARD II data, which was also the second best performing system in the DIHARD II challenge \cite{lin2020dihard}. It is intriguing to note that our embeddings perform well in the meeting domain despite the presence of high overlapping speech error. This is possible because the proposed embeddings were trained on meeting data. However, x-vectors that are extracted from Voxceleb trained model also perform well. The other domains where we obtain noticeable improvements over the existing system include court, child, socio\_field, and socio\_lab.   

\vspace{-12pt}
\subsection{ADOS and BOSCC databases}

Finally, we evaluate the proposed method on two child-clinician interaction corpora from the domain of Autism Spectrum Disorder: ADOS and BOSCC. 
The diarization results are presented in Table~\ref{table7}. We observe from the table that the Kaldi x-vector diarization system (last row) does not perform well on these two databases. The most probable reason behind this is that the PLDA model is trained on Voxceleb data and thus creating a significant domain mismatch. However, the x-vectors with k-means and SC perform reasonably well on both ADOS and BOSCC data than the Kaldi x-vector system. Among the proposed embeddings, ClusterGAN is superior to MCGAN both individually and also when fused with x-vectors. This is attributed to the better performance of ClusterGAN over MCGAN in the known number of speakers condition in general. A significant reduction in DER is seen while SC is employed as the clustering mechanism. The best achieved DER on ADOS and BOSCC datasets is 6.74\% and 9.26\%, respectively, and this is for the x-vector + ClusterGAN with SC system. We obtain a relative improvement of 53.06\% and 57.31\% over Kaldi x-vector on the ADOS and BOSCC databases respectively. We note that although we expect better generalization from MCGAN due to meta-learning, ClusterGAN emerges as useful in these known number of (dyadic) speaker conditions, i.e., child and adult interlocutors.

\vspace{-12pt}
\section{Conclusions}\label{section7}

We proposed new speaker embeddings by exploiting the latent space of GANs using ClusterGAN and by making the encoder in the ClusterGAN more robust and generalizable with the help of prototypical loss fine-tuning. We benchmarked the proposed embeddings individually and also fused with x-vectors within the speaker diarization framework. 
We investigated the effectiveness of the proposed embeddings by extensively evaluating them for speaker diarization across five different databases. 
We obtain a relative improvement of 19.43\%, 72.32\%, 30.64\%, 53.06\%, and 57.31\% over the Kaldi x-vector baseline on CALLHOME, AMI-eval, DIHARD II dev, ADOS, and BOSCC databases respectively. The key findings of this work can be summarized as follows:
\begin{itemize}
    \item MCGAN embeddings outperform x-vectors and ClusterGAN embeddings significantly on telephonic data for both known and automatically estimated number of speaker conditions with both k-means and SC back-end. They also perform better than ClusterGAN in the estimated number of speaker condition on meeting and multi-domain datasets.
    \item Analysis suggests that MCGAN embeddings are robust in the number of speakers estimation and in diarizing sessions which have significant presence of short speech segments when compared to x-vectors, ClusterGAN and fused embeddings. 
    \item Embedding fusion of x-vectors and the proposed embeddings improves diarization performance consistently for all the corpora considered. Therefore, we speculate that both the proposed embeddings have complementary information to the x-vectors. The proposed fused embeddings with NME-SC outperform the Kaldi x-vector system and emerges as the top-performing system on the challenging multi-domain DIHARD II dev set.
    
    
\end{itemize}
In the future, it would be worthwhile to investigate speech spectrograms directly instead of pre-trained embeddings as the input. The usage of other existing meta-learning algorithms will also be explored in the context of speaker diarization.

\begin{table}[!t]
\caption{Results on ADOS and BOSCC databases for the baseline and proposed systems.}
\vspace{-8pt}
\centering
\setlength{\tabcolsep}{3pt}
\begin{tabular}{cccc}
\Xhline{2.5\arrayrulewidth}
Embedding   & Back-end     & \begin{tabular}[c]{@{}c@{}}Avg. DER (\%) \\ (oracle SAD) \\ on ADOS\end{tabular} & \begin{tabular}[c]{@{}c@{}}Avg. DER (\%) \\ (oracle SAD) \\ on BOSCC\end{tabular} \\
\Xhline{2.5\arrayrulewidth}

x-vector   & \multirow{5}{*}{k-means}    & 12.35                         & 14.73                            \\
ClusterGAN  &          & 10.21                         & 10.59                            \\
MCGAN    &        & 9.71                         & 14.67                            \\
x-vector + ClusterGAN &  & 8.70                         & 10.52                            \\
x-vector + MCGAN & & 9.10                          & 13.22                           \\ \hline
x-vector  &\multirow{5}{*}{SC}          & 8.51                        & 11.99                           \\
ClusterGAN  &         & 6.75                        & 9.32                            \\
MCGAN    &        & 9.96                         & 13.21                            \\
x-vector + ClusterGAN &  & \textbf{6.74}                        & \textbf{9.26}                            \\
x-vector + MCGAN & & 9.18                          & 12.17                           \\
\Xhline{2.5\arrayrulewidth}
Kaldi x-vector   &PLDA+AHC   & 14.36                         & 21.69                            \\\Xhline{2.5\arrayrulewidth}
\end{tabular}\label{table7}
\vspace{-12pt}
\end{table}


%





\ifCLASSOPTIONcaptionsoff
  \newpage
\fi



%



\bibliographystyle{IEEEtran}
\vspace{-10pt}
\bibliography{main}

%








\end{document}